\documentclass[aps,prb,reprint,superscriptaddress,showpacs]{revtex4-1}

\usepackage{graphicx}
\usepackage{dcolumn}
\usepackage{bm}
\usepackage{setspace}
\usepackage{amsfonts}

\newcommand{\dmdt}{\ensuremath{\mathrm{d}m/\mathrm{d}t}}

\newcommand{\wcrit}{\ensuremath{w_{\mathrm{c}}}}

\renewcommand{\v}{\mathbf{v}}

\newcommand{\dn}{d_{\mathrm{notch}}}
\newcommand{\wn}{w_{\mathrm{notch}}}
\newcommand{\ft}{f_{\mathrm{trans}}}
\newcommand{\fb}{f_{\mathrm{breathe}}}
\newcommand{\ftw}{f_{\mathrm{twist}}}
\newcommand{\lmesh}{l_{\mathrm{mesh}}}
\newcommand{\hdepin}{H_{\mathrm{depin}}}

\newcommand{\xtdw}{x_{\mathrm{TDW}}}

\begin{document}

\title{Resonant translational, breathing and twisting modes of pinned transverse magnetic domain walls}

\author{Peter J.~Metaxas}
\email{peter.metaxas@uwa.edu.au}
\affiliation{
School of Physics, M013, University of Western Australia, 35 Stirling Hwy, Crawley WA 6009, Australia.}%
\affiliation{Unit\'e Mixte de Physique CNRS/Thales and Universit\'e Paris-Sud, 1 Avenue A.~Fresnel, Palaiseau, France}%

\author{Maximilian~Albert}
\affiliation{Engineering and the Environment, University of Southampton, Southampton, United Kingdom}

\author{Steven~Lequeux} 
\author{Vincent~Cros}
\author{Julie~Grollier}
\author{Paolo~Bortolotti}
\author{Abdelmadjid~Anane}
\affiliation{Unit\'e Mixte de Physique CNRS/Thales and Universit\'e Paris-Sud, 1 Avenue A.~Fresnel, Palaiseau, France}

\author{Hans Fangohr}
\affiliation{Engineering and the Environment, University of Southampton, Southampton, United Kingdom}

\begin{abstract}We study translational, breathing and twisting resonant modes of transverse magnetic domain walls pinned at notches in ferromagnetic nanostrips. We demonstrate that a mode's sensitivity to notches depends strongly on the characteristics of that particular resonance. For example, the frequencies of modes involving lateral motion of the wall are the ones which are most sensitive to changes in the notch intrusion depth (especially at the narrower, more strongly confined end of the domain wall). In contrast, the breathing mode, whose dynamics are concentrated away from the notches is relatively insensitive to changes in the notches' sizes. We also demonstrate a sharp drop in the translational mode's frequency towards zero when approaching depinning which is found, using a harmonic oscillator model, to be consistent with a reduction in the local slope of the notch-induced confining potential at its edge.
\end{abstract}


\maketitle

\section{Introduction}

Domain walls (DWs) separate oppositely oriented magnetic  domains in ferromagnetic strips and have applications  ranging from data storage\cite{Parkin2008} to neuromorphic computing \cite{Wang2009,Locatelli2013} and biotechnology \cite{Donolato2010,Rapoport2012a}. Applications typically exploit DW displacement and/or resonant DW excitations\cite{Saitoh2004}, the latter corresponding to precessional magnetization dynamics localized at the DW\cite{Winter1961,Saitoh2004,Rebei2006,Bedau2007,Sandweg2008,Roy2010,Sangiao2014,Lequeux2015}. These excitations can can be exploited in oscillators \cite{Lepadatu2010} and magnonic devices\cite{Bayer2005,Hermsdoerfer2009} as well as for assisting  domain wall motion\cite{LeMaho2009,Han2009,Jamali2010,Janutka2013,Wang2014} or depinning\cite{Bedau2007,Thomas2007,Nozaki2007,Martinez2008,Metaxas2010apl}. 

\begin{figure*}
\centering
	\includegraphics[width=15cm]{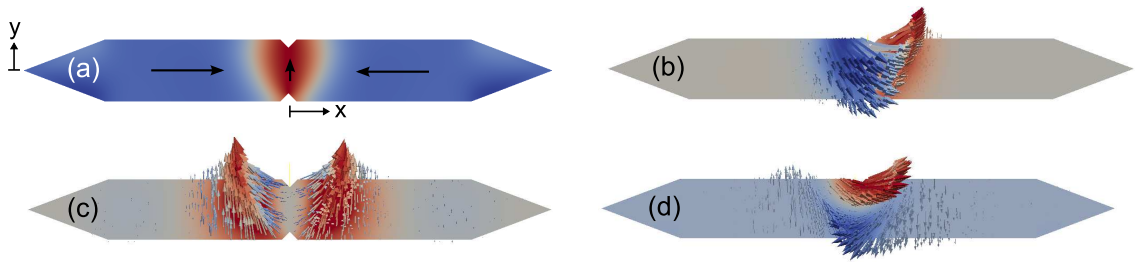}
	\caption{(Color online) (a) Zero-field equilibrium magnetization configuration, $\mathbf{m}_0(\mathbf{r})$, in a 75~nm wide NiFe strip with symmetric notches ($\wn=20$~nm, $\dn=10$~nm) containing a head-to-head TDW with $m_y$ color scaling. The black arrows indicate the local magnetization direction. The $x$ and $y$ axis origins are also shown. (b-d) Snapshots of the translational, breathing and twisting modes  showing the dynamic component only ($\mathbf{dm}(\mathbf{r})$). The  translational mode snapshot (b) uses  $m_y$ color scaling and is taken when the TDW is displaced to the right ($+x$) at which point there is a significant dynamic $+m_x$ component.  The  breathing mode snapshot (c) also uses  $m_y$ color scaling and is taken at the point during the TDW width oscillation when the  width is larger than its equilibrium value. There is thus a large dynamic $+m_y$ component at the TDW edges which broadens the TDW.  The  twisting mode snapshot (d) uses $m_x$ color scaling and is taken at the point when the wide end of the TDW ($+y$) is displaced to the right and the narrow end of the TDW ($-y$) is displaced to the left. See also animations of the modes\cite{Note1}.}
	\label{fig:modes}
\end{figure*}

The use of resonant phenomena in applications will however rely on successful control of the resonant modes of DWs. It is known that large geometrical constrictions such as notches (also widely used for positional control \cite{Parkin2008,Petit2008,Bogart2009,Kunz2010}) in micron-scale strips can  be  used to tune the frequency of a DW's translational mode \cite{Lepadatu2010}. For smaller\cite{Currivan2014} device geometries however, uniform fabrication of small  notches may become challenging since  the notches' dimensions will likely become comparable to those of lithographic  defects and edge roughness. 

In this work we show how different DW resonances have different sensitivities to notches and that these sensitivities can be linked to the nature of the mode and the structure of the DW. For example, modes which involve either local or global translation of the wall can be highly sensitive to the presence, size and position of the notch. Our work focuses on   the resonant properties of pinned head-to-head transverse domain walls [TDWs, Fig.~\ref{fig:modes}(a)] which arise in thin, narrow, in-plane magnetized strips \cite{Nakatani2005}. The TDWs are pinned at triangular notches located at the edge of the strip. 
We use a numerical eigenmode method to study  three TDW resonances, corresponding to translational\cite{Rebei2006,Bedau2007,Lepadatu2010,Rhensius2010}, twisting\cite{Roy2010,Wang2013b} and breathing \cite{Stamps1997,Liu1998,Dantas2001,Rebei2006,Matsushita2012,Mori2014} excitations of the TDW. The latter mode has recently been  studied  for oscillator applications \cite{Matsushita2014} and we demonstrate that this mode has the lowest sensitivity to changes in notch depths, making it an appealing choice for device applications.
The eigenmode method we use also enables us to study the modes in the vicinity of the static depinning field where we find a sharp drop off in the translational mode frequency. This dramatic change in frequency can be linked directly to the position-dependence of the slope of the notch-induced confining potential (probed here by field-induced displacements of the TDW within the potential).

\section{Micromagnetic simulation method}

Many numerical studies of resonant modes in confined geometries use time domain (`ringdown') methods in which Fourier analysis of precessional magnetization dynamics is employed to extract resonant mode frequencies and spatial profiles. These methods require the system to be subjected to an external excitation \cite{Roy2010,Grimsditch2004,McMichael2005,Dvornik2011,Wang2013b}, often a pulsed magnetic field. In contrast, eigenmode methods \cite{dAquino2009,Naletov2011} enable a direct calculation of  resonant magnetic modes from a system's equilibrium magnetic configuration, $\mathbf{m}_0(\mathbf{r})$ (as do  dynamical matrix methods \cite{Zivieri2012a}).  This enables the observation of the full mode spectrum without requiring careful choice of the  ringdown excitation's symmetry. It also enables us to study DW modes at fields which are in the close neighborhood of the static depinning field where excited translational resonances could otherwise
resonantly depin\cite{Bedau2007,Thomas2007,Nozaki2007,Martinez2008,Metaxas2010apl} the wall.

Our  simulations were run on a  Permalloy strip having saturation magnetization $M_S=860$ kA/m and exchange stiffness $A_\mathrm{ex}=13$ pJ/m using the finite element micromagnetic package Finmag (successor to Nmag \cite{Fischbacher2007}). The strip has tapered ends 
and two central notches for TDW pinning  [Fig.~\ref{fig:modes}(a)]. Unless otherwise noted, the notches are  located at $x=0$, the strip thickness is 5 nm and the total length is 750 nm.

Magnetic eigenmodes are determined from $\mathbf{m}_0(\mathbf{r})$ using a method which is similar to that described by d'Aquino \textit{et. al.}\cite{dAquino2009}~and which is  valid for small time-dependent  oscillations $\mathbf{dm}(\mathbf{r},t)$ around $\mathbf{m}_0(\mathbf{r})$. The basic principle is to linearize the LLG equation around the equilibrium state $\mathbf{m}_0(\mathbf{r})$, resulting in a linear system of ordinary differential equations (ODEs) for the oscillations $\mathbf{dm}(\mathbf{r},t)$. This system of ODEs can be phrased as an eigenvalue problem for  $\mathbf{dm}(\mathbf{r},t)$ which has a full set of solutions representing the eigenmodes of the nanostrip. The complex coefficients of each solution vector encode the local amplitudes and relative phases of the eigenmode at the nodes of the finite element mesh. The  algorithm computes a number of modes of increasing frequency, $f$. Each $f$ has a real and an imaginary part with the latter typically 3-4 orders of magnitude smaller than the former. We quote the real parts of $f$. In theory, the eigenfrequencies are purely real; the small imaginary part stems from the formulation of the problem as a non-Hermitian eigenvalue problem for which the eigensolver returns complex solutions with a small imaginary component due to numerical inaccuracies. Eigenmodes localized at the TDW can be identified by visual inspection of the spatially resolved eigenvectors. Either the dynamic component, $\mathbf{dm}(\mathbf{r},t)$, may be inspected alone or it can be scaled and added to $\mathbf{m}_0(\mathbf{r})$, enabling a visualization of the actual TDW dynamics for each mode (e.g.~see mode animations\footnote{We intend to include supplementary animations with the final published version of the paper.}).

To find $\mathbf{m}_0(\mathbf{r})$, the system was initialized with a trial head-to-head TDW configuration centered on $x=0$ and allowed to relax with damping parameter $\alpha=1$, typically until $\dmdt < 1^{\circ}/$ns at all points in the strip. For a strip width of 75~nm and a thickness of 5~nm, using the stricter criterion $\dmdt < 0.1^{\circ}/$ns resulted in changes in the mode frequencies of  1.1 Mhz or less ($\leq 0.04$\%). The relaxed configuration was a pinned TDW for all studied geometries \cite{Nakatani2005}. Note that the TDW [Fig.~\ref{fig:modes}(a)] is wider at the $+y$ side of the strip which will be important for determining TDW-notch interactions.

We used a characteristic internode length for the finite element mesh of $\lmesh=3$~nm at $x=0$ (less than the NiFe exchange length of 5.7 nm \cite{Abo2013}) with a smooth transition to a larger $\lmesh=8$ nm at the ends of the strip to reduce computational time and memory use. As such, we present results only on those modes which are localized on the TDW near the center of the strip). We note that except for those simulations in which magnetic fields close to the DW depinning field are applied, the error in the mode frequency associated with the larger $\lmesh$ at the device ends was found to be less than 1\%.  A post-relaxation mesh coarsening\cite{Metaxas2015} could potentially be applied to future studies. Finally, we note that a comparison with a time domain ringdown simulation is given in Appendix \ref{ringdown}.

\begin{figure*}[hbt]
	\includegraphics[width=11cm]{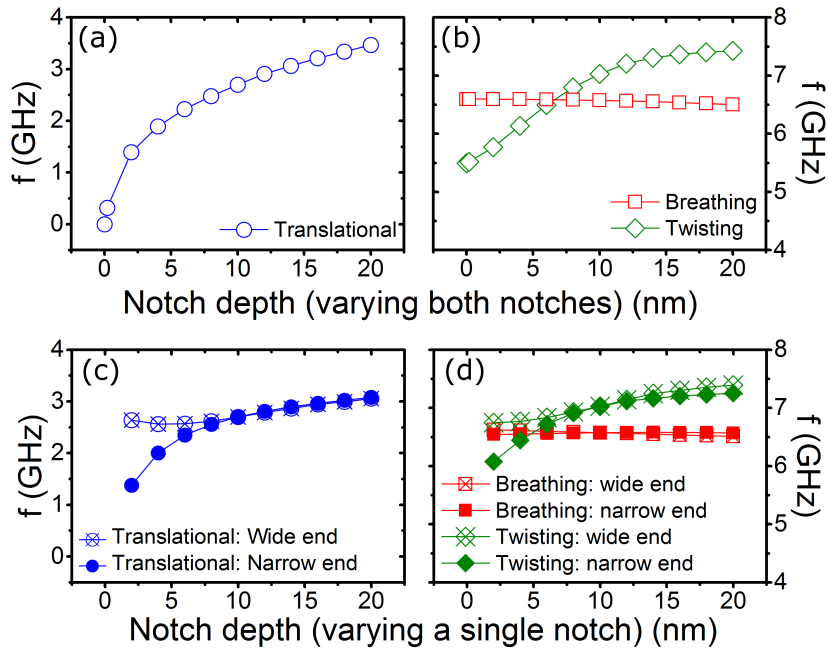}
	\caption{(Color online) (a,b) TDW eigenfrequencies versus $\dn$ when varying $\dn$ for both notches simultaneously. (c,d) Eigenfrequencies when varying $\dn$ only at one side of the strip, either at the wide end or narrow end of the wall while keeping the other notch with $\dn=10$ nm. For all data $\wn=20$.}
	\label{f:notch}
\end{figure*}

\section{TDW modes}

The three lowest frequency TDW modes  correspond to translational, breathing or twisting deformations. In Figs.~\ref{fig:modes}(b-d) these three calculated modes are shown  (as a snapshot of the mode's dynamic component, $\mathbf{dm}(\mathbf{r})$ at a time such that $\mathbf{dm}(\mathbf{r})$ is large) for a 75~nm strip with symmetric, triangular notches, each with width, $\wn=20$ nm and a depth of intrusion into the strip, $\dn=10$~nm. The translational mode (2.70 GHz) corresponds to an oscillatory, side-to-side motion of the TDW away from the notches  [Fig.~\ref{fig:modes}(b)]. 
For the breathing mode\cite{Stamps1997,Liu1998,Dantas2001,Rebei2006,Matsushita2012,Wang2013b,Mori2014} [6.57~GHz, Fig.~\ref{fig:modes}(c)], dynamics are concentrated at the edges of the domain wall with the excitations mirrored around $x=0$. This leads to an oscillatory change in the TDW's width as a function of time. For this strip width, the highest frequency mode is the 7.03 GHz twisting mode [Fig.~\ref{fig:modes}(d)] which involves the TDW's two ends (near the top/bottom of the strip) moving in opposite directions. Idealizing the TDW as a  string crossing the nanostrip, this  mode has similarities to a standing wave with a zero-displacement node ($\mathbf{dm}\approx 0$) near $y=0$. Wang \textit{et al.}\cite{Wang2013b} have observed what seem to be similar breathing and twisting modes of unpinned TDWs. As shown below, and in contrast to what is observed for the translational mode, a finite frequency for the breathing and twisting modes is non-reliant on confinement (i.e.~they are intrinsic $f>0$ TDW excitations).

\subsection{Notch dependence}

The translational and twisting  modes both  involve some movement of the TDW away from the energetically favorable $x=0$ position. This can either be a global side-to-side movement of the TDW (as for the translational mode) or a local side-to-side movement (as for the twisting mode where out of phase lateral TDW movements arise at opposite edges of the strip). This has strong implications for notch  sensitivity with the twisting and translational modes having a strong dependence on the notch size. In contrast, dynamics of the breathing mode are concentrated away from the notch at the TDW's edges which results in a much weaker sensitivity to the notch size. 

To demonstrate this, we have plotted each TDW eigenfrequency in Figs.~\ref{f:notch}(a,b)  as a function of the notches' intrusion depths for a 75 nm wide strip with a $20$ nm ($=\wn$) wide notch and with both notches having the same geometry on each side of the strip. One will notice immediately that the twisting and translational modes (i.e.~those with some translational nature) are highly dependent on $\dn$. The translational mode's frequency, $\ft$, decreases smoothly with $\dn$, going to zero at $\dn=0$ Fig.~\ref{f:notch}(a). This is consistent with the wall being free to translate laterally at $\ft=0$ in the absence of pinning (i.e.~$\dn=0$ corresponds to a smooth-edged strip with no notches).  The twisting mode frequency, $\ftw$, also depends quite strongly on $\dn$,  reducing by $\sim$40\% ($\sim 2$ GHz) when changing $\dn$ from 20 nm to 0 nm [Fig.~\ref{f:notch}(b)]. In contrast, the breathing mode frequency, $\fb$, changes by only 1.5\% over the same range of $\dn$ values [Fig.~\ref{f:notch}(b)].  Note also that $\fb$ and $\ftw$ remain finite at $\dn=0$, consistent with these modes being intrinsic TDW  excitations for which the observation of a finite eigenfrequency is non-reliant on notch-induced, lateral TDW confinement.

\begin{figure}
	\includegraphics[width=8.5cm]{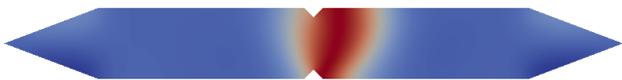}
	\caption{(Color online) Deformed domain wall in a 75 nm strip for $H_x=5530$ A/m. }
	\label{f:shiftedwall}
\end{figure}

Despite both notches being geometrically identical, one can see from the mode snapshots in  Figs.~\ref{fig:modes}(b,d) that both the twisting and translational modes' dynamics are largest at the wide end of the TDW. This suggests that this end of the TDW has a weaker lateral  confinement than the narrow end of the TDW. This is confirmed in  Fig.~\ref{f:shiftedwall} which shows a TDW being pushed away from the notches under the action of a magnetic field, $H$, applied along the $x$ axis ($H<\hdepin$, the static  depinning field). It is the less strongly pinned wide end of the TDW which is displaced furthest from the notch. To see what effect each notch has on the modes,  we show in Figs.~\ref{f:notch}(c,d) results obtained while varying $\dn$  on only one side of the strip (either at the wide end or at the narrow end of the TDW) while keeping the other notch's intrusion depth fixed at 10 nm. We indeed find that $\ft$ is most sensitive to changes of $\dn$ at the narrow end of the wall, that notch being dominant in determining $\ft$.  Reducing $\dn$ from 10~nm to 2~nm at the narrow end of the wall [filled circles in Fig.~\ref{f:notch}(c)] generates a 40\% reduction in  $\ft$.  Notably, this is accompanied by a transition to a more pure translation of the TDW structure in its entirety rather than an excitation in which the highest amplitude dynamics occur at the wide end of the TDW [as in Fig.~\ref{fig:modes}(a)]. In contrast,  $\ft$ remains fairly constant when changing $\dn$ only at the wide end of the wall  [crossed open circles in Fig.~\ref{f:notch}(c)]. This trend also holds for $\ftw$. The $\dn$-dependence of $\fb$ remains weak.

\begin{figure}
	\includegraphics[width=6.5cm]{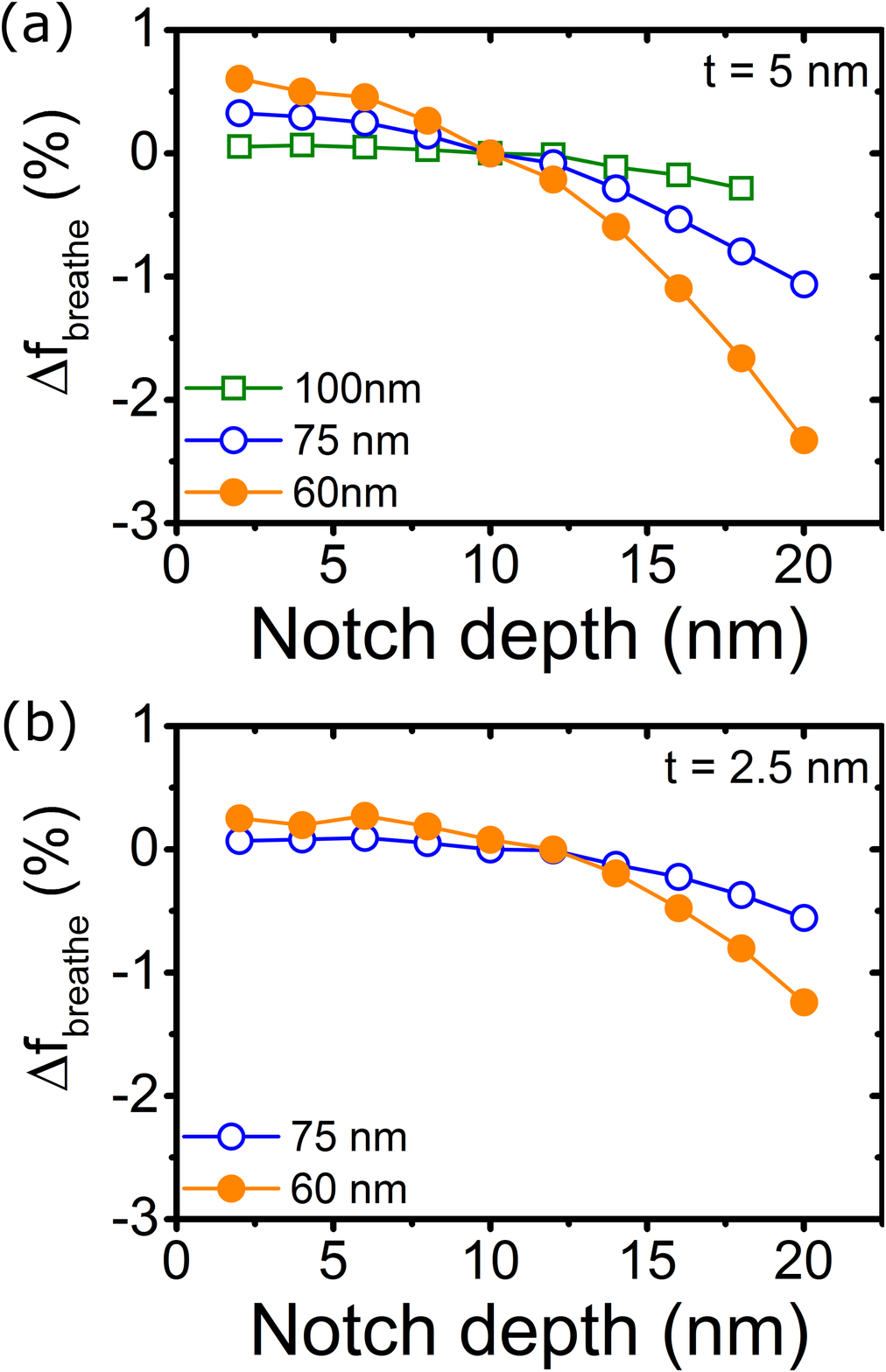}
	\caption{(Color online) Percentage change in $\fb$ with respect to $\fb$ at $\dn=10$  plotted against $\dn$ for (a) 5 nm thick strips and (b) 2.5 nm thick strips at various strip widths (see legends).}
	\label{f:checkbreathe}
\end{figure}

To test the limits of the $\dn$-insensitivity of $\fb$, simulations were run with the notch at the wide end of the wall displaced away from $x=0$ for the 75 nm wide strip. This did lead to small changes in $\fb$  ($\dn=10$ nm, $\wn=20$ nm) with some distortion of the breathing mode observed  when the notch was right at the edge of the TDW. However the maximum frequency change still remained within $3$\% of the value observed for two laterally centered notches. We also looked at the percentage variation of $\fb$ for two other strip widths (60 nm and 100 nm wide 5 nm thick strips) for centrally located notches. We found the lowest sensitivity for larger widths where the notch intrudes comparatively less far into the strip and thus presumably generates the weakest change to the energy landscape that is experienced by the TDW. Reducing the thickness of the layer also led to a reduced sensitivity. This can be seen in Fig.~\ref{f:checkbreathe}(b) where we again plot resonance data for 60 nm and 75 nm wide strips but this time with a reduced (2.5 nm) strip thickness. An important point to note from  Fig.~\ref{f:checkbreathe} is that the breathing mode remains highly insensitive to changes in $\dn$ for  small notches at all studied widths. We  see the alrgest changes in $\fb$ when $\dn$ becomes larger than about 12 nm suggesting that small defects should have only a minor effect on the breathing mode. In contrast, the other two modes are least sensitive to changes in $\dn$ when $\dn$ is large (Fig.~\ref{f:notch}).

We briefly note that changes in the \textit{width} of the notch (for a fixed notch depth of 10 nm) yielded  weak changes for both $\fb$ and $\ftw$. Over a range of notch widths from 5 nm to 50 nm we observed  $\Delta \ftw \leq 3$ \% and $\Delta \fb \leq 2$ \%. The change in $\ft$ was also minor when reducing the notch width below 20 nm ($\Delta \ft \leq 6$ \%). However, broadening the notch  to 50 nm led to a strong reduction in $\ft$ of $>60$ \%, presumably due to a strongly reduced confinement by the broader notches (the effect of confinement on $\ft$ is discussed further below).

\subsection{Strip width dependence}\label{widthsection}

\begin{figure*}
\includegraphics[width=11cm]{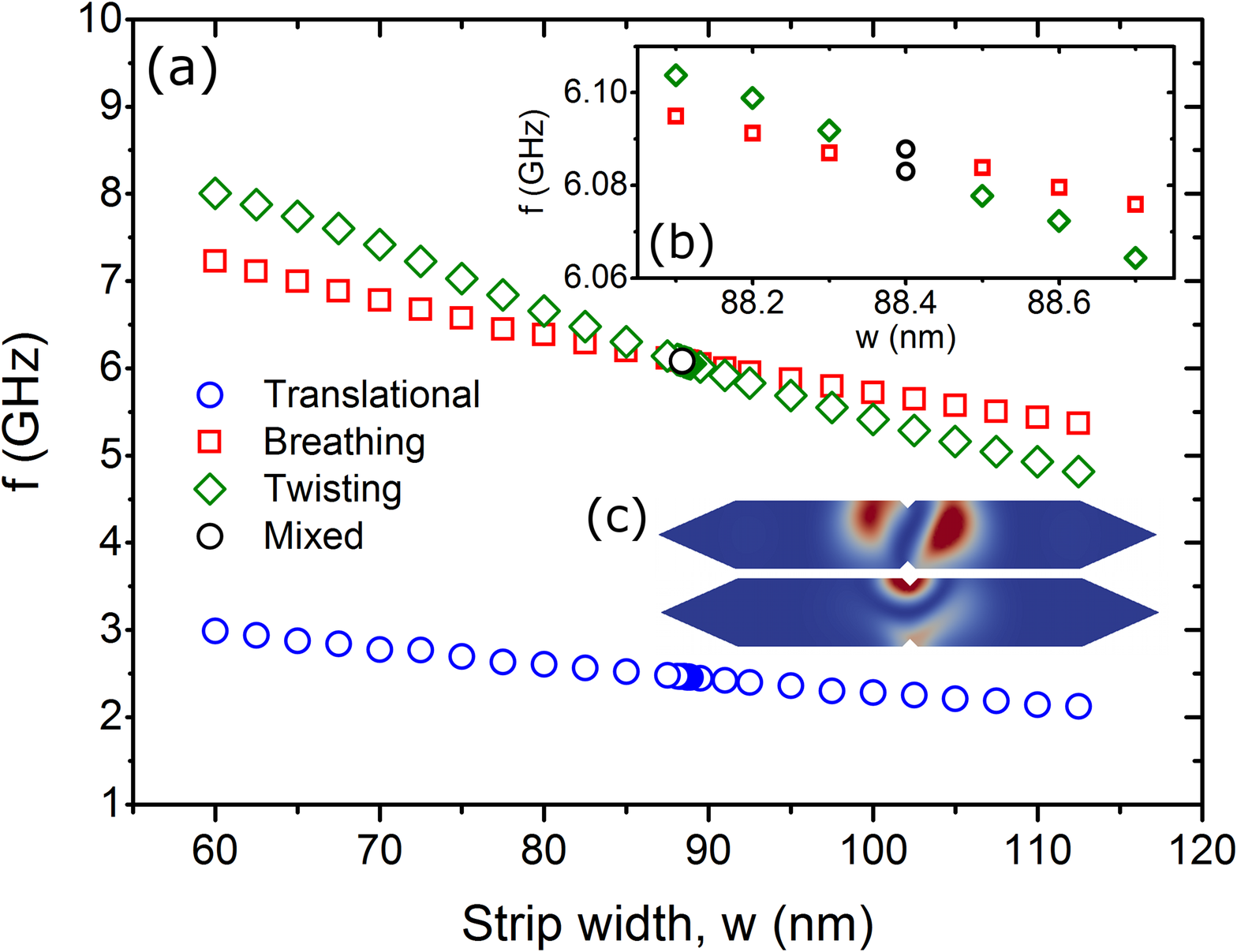}
\caption{(Color online) (a) Frequencies of the three TDW eigenmodes as a function of strip width, $w$. The notches are symmetric ($\dn=10$ nm, $\wn=20$ nm). At $w=88.4$~nm the calculated modes are mixed breathing-twisting modes (see inset, b). (c) shows snapshots of the amplitude of the dynamic component (red) of  the mixed modes found for $w=88.2$~nm at 6.091~GHz (upper, primarily a breathing mode) and 6.099~GHz (lower, primarily a twisting mode).}
\label{f:widths}
\end{figure*}

When holding the notch geometry constant ($\wn =20$ nm and $\dn =10$ nm), we find that increasing the strip width leads to an reduction in each of the TDW mode's frequencies 
[Fig.~\ref{f:widths}(a)]. The breathing and twisting modes remain highest in frequency and their similar frequencies, coupled with slightly different width dependencies, results in a mode crossing which occurs at $w=\wcrit\approx 88.4$ nm for this 5 nm thick strip [Figs.~\ref{f:widths}(b)]. At $w\approx \wcrit$, a translational mode as well as two other distinct TDW modes are found with the latter appearing as mixed twisting-breathing modes [e.g.~Fig.~\ref{f:widths}(c)]. Analogous mixed modes were also calculated for a similar geometry using the mode solver in the SpinFlow3D simulation package (some details on this  solver have been given previously  \cite{Naletov2011}).  This mode mixing can be clearly identified when visually inspecting the modes for $|w-\wcrit|\lesssim 1.5$~nm. As $|w-\wcrit|$ increases, the computed  modes become more `pure' (i.e.~a dominant breathing or twisting characteristic). In  Fig.~\ref{f:widths}(b),  all modes at $w\ne 88.4$~nm are labeled either as twisting or breathing with the label corresponding to the mode which is dominant. Note that we expect no mode coupling in this eigenmode approach since this would require the inclusion of damping in the mode determination \cite{dAquino2009}. Indeed, we have found that the mixing arises due to the arbitrary basis chosen by the eigensolver: each mixed mode  is a linear combination of the `pure' orthogonal twisting and breathing eigenmodes (Appendix \ref{mixing}).

\subsection{Width dependent confinement and its effect on the translational mode}

\begin{figure}
	\includegraphics[width=7cm]{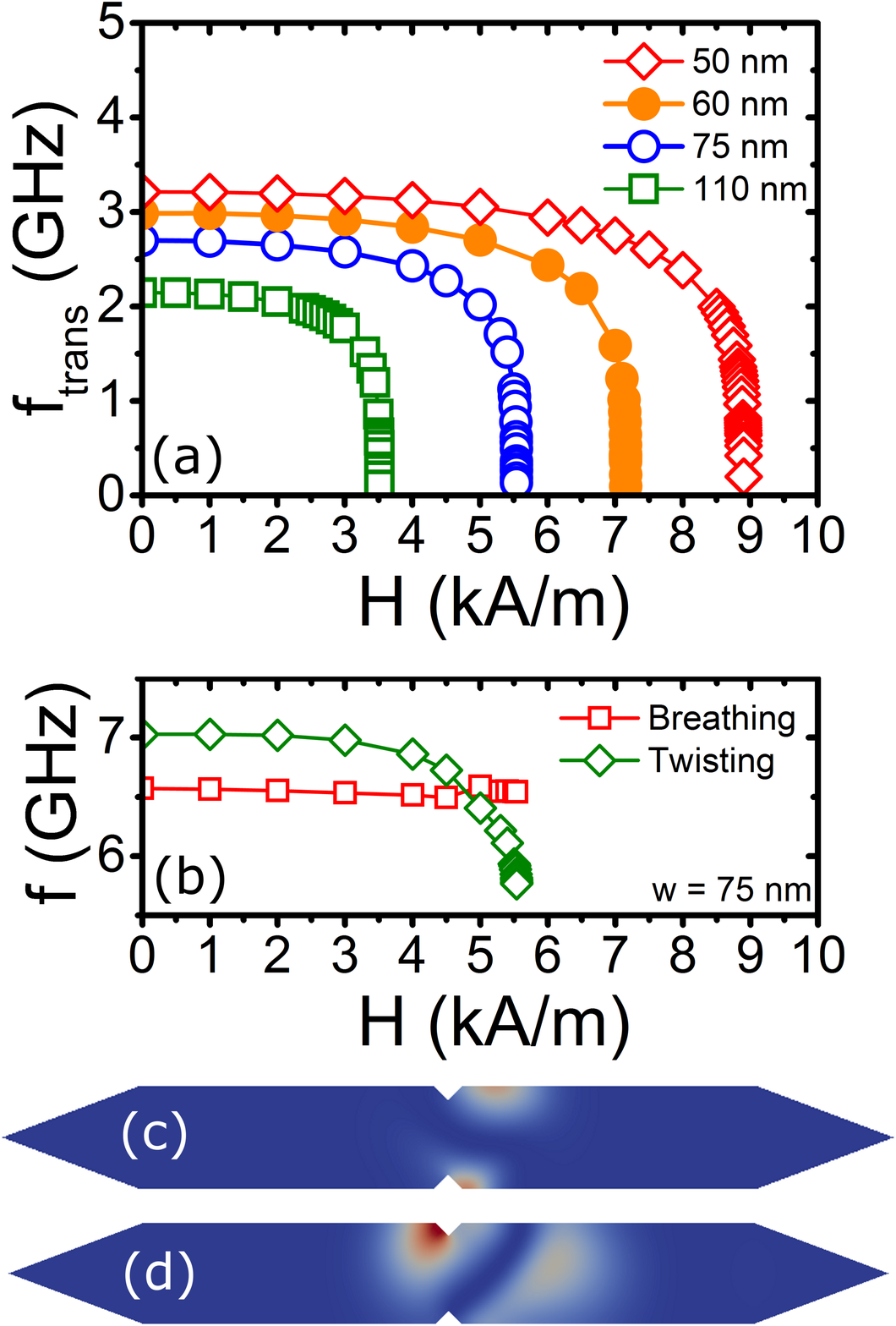}
	\caption{(Color online) (a) $\ft$ versus in-plane field, $H$ (oriented along $+x$), for strip widths of 50, 60, 75 and 110 nm ($\dn=10$ nm and $\wn=20$ nm). (b) $\fb$ and $\ftw$ versus $H$ at a strip width of 75 nm. (c,d) Snapshots of the amplitude of the dynamic component (red) of the magnetization for the (c)  twisting and (d) breathing modes at a strip width of 75 nm for $H=5530$ A/m (i.e.~close to depinning).}
	\label{f:depinning}
\end{figure}

We now turn to the width dependence of the translational mode which will be shown to be linked to the width-dependence of the notch-induced confinement of the TDW. Note that some  qualitative models for the higher frequency breathing and twisting mode frequencies as a function of strip width are given in Appendix \ref{othermodes}.  The  frequency of the translational mode of the pinned TDW, $\ft$, as a function of $H< \hdepin$ is shown for  a number of strip widths in Fig.~\ref{f:depinning} (again we use $\wn =20$ nm and $\dn =10$ nm). Note that for $H>\hdepin$, $\mathbf{m}_0(\mathbf{r})$ is that of a quasi-uniformly magnetized strip with no DW. As such, there is no translational mode frequency data above $\hdepin$. For all strip widths, $\ft$ shows a weak negative monotonic dependence on $H$ for small $H/\hdepin$. However, $\ft$ drops sharply to zero (i.e.~again going toward the case of a free TDW) as $H\rightarrow \hdepin$. Note that for $H\approx\hdepin$, $\ft$ exhibits a stronger sensitivity to the relaxation parameters of the simulation, requiring the use of a smaller $\dmdt$ near $\hdepin$. $\ft$ as well as the determined value of $\hdepin$ itself is also  more sensitive to the non-uniform meshing than the undeformed TDW at $H=0$. For example,  a slightly higher $\hdepin$ ($<1$\% relative change) was found when using $\lmesh=3$ nm throughout the structure at $w=60$ nm. Thus, there is some influence of the meshing on the pinning of the wall here.

\begin{figure*}
	\includegraphics[width=13cm]{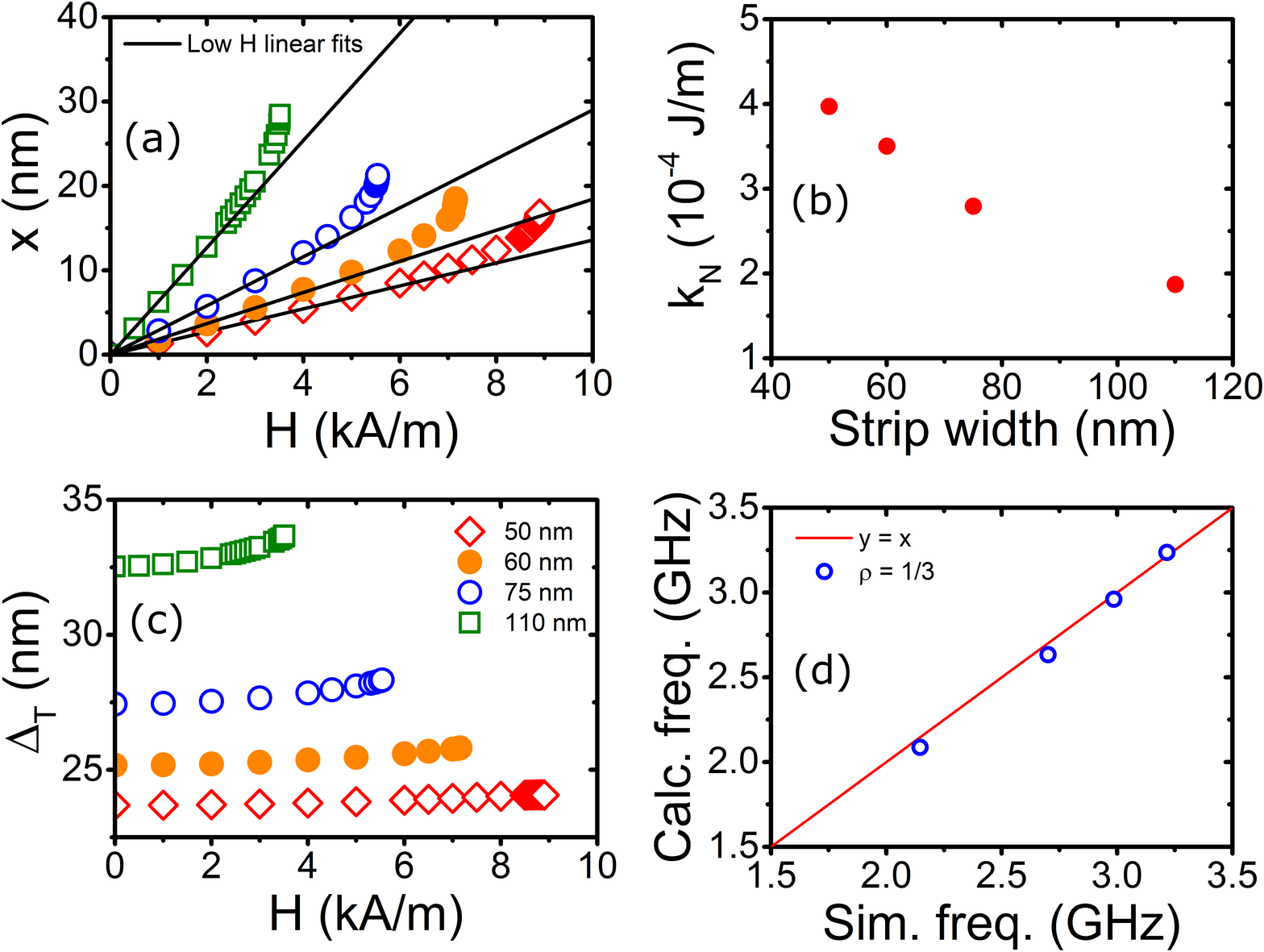}
	\caption{(Color online) (a) TDW position versus $H$ applied along the $+x$ direction. Solid lines are linear fits to the low field data (typically the first 4-5 points). (b) Calculated TDW spring constant versus strip width. (c) Thiele domain wall width of the $H$-deformed TDWs versus $H$. (d) Calculated $\ft$ (calculated as per the text using the data in (a,c) and Eqs.~(1-3)) versus the simulated $\ft$ taken from the data in Fig.~\ref{f:depinning}.}
	\label{f:pinpot}
\end{figure*}

\begin{figure}
	\includegraphics[width=6cm]{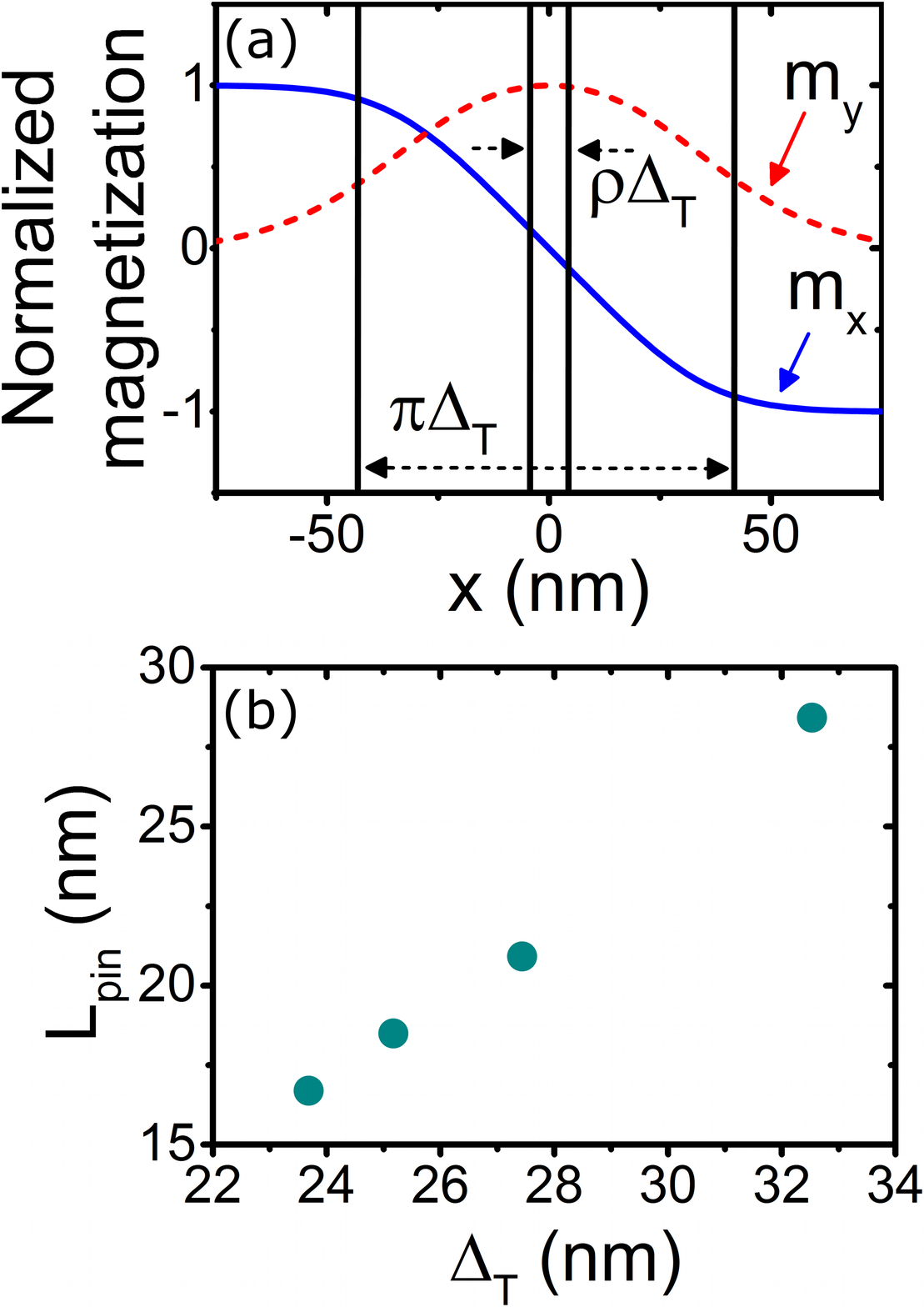}
	\caption{(Color online) (a) $x$-dependence of the $x$ and $y$ components of the magnetization taken at $y=0$ (at the center of the strip). $\Delta_T$ is the Thiele DW width and $\rho$ is a scaling factor used in the demagnetizing field calculation. (b) Effective width of the pinning potential ($L_{\mathrm{pin}}$)  estimated from the maximum displacement of the TDW before depinning (taken from Fig.~\ref{f:pinpot}(a)) plotted against $\Delta_T$ for strip widths of 50, 60, 75 and 110 nm. The largest width strip has the largest $\Delta_T$. }
	\label{f:lpin}
\end{figure}

\begin{figure*}
	\includegraphics[width=12cm]{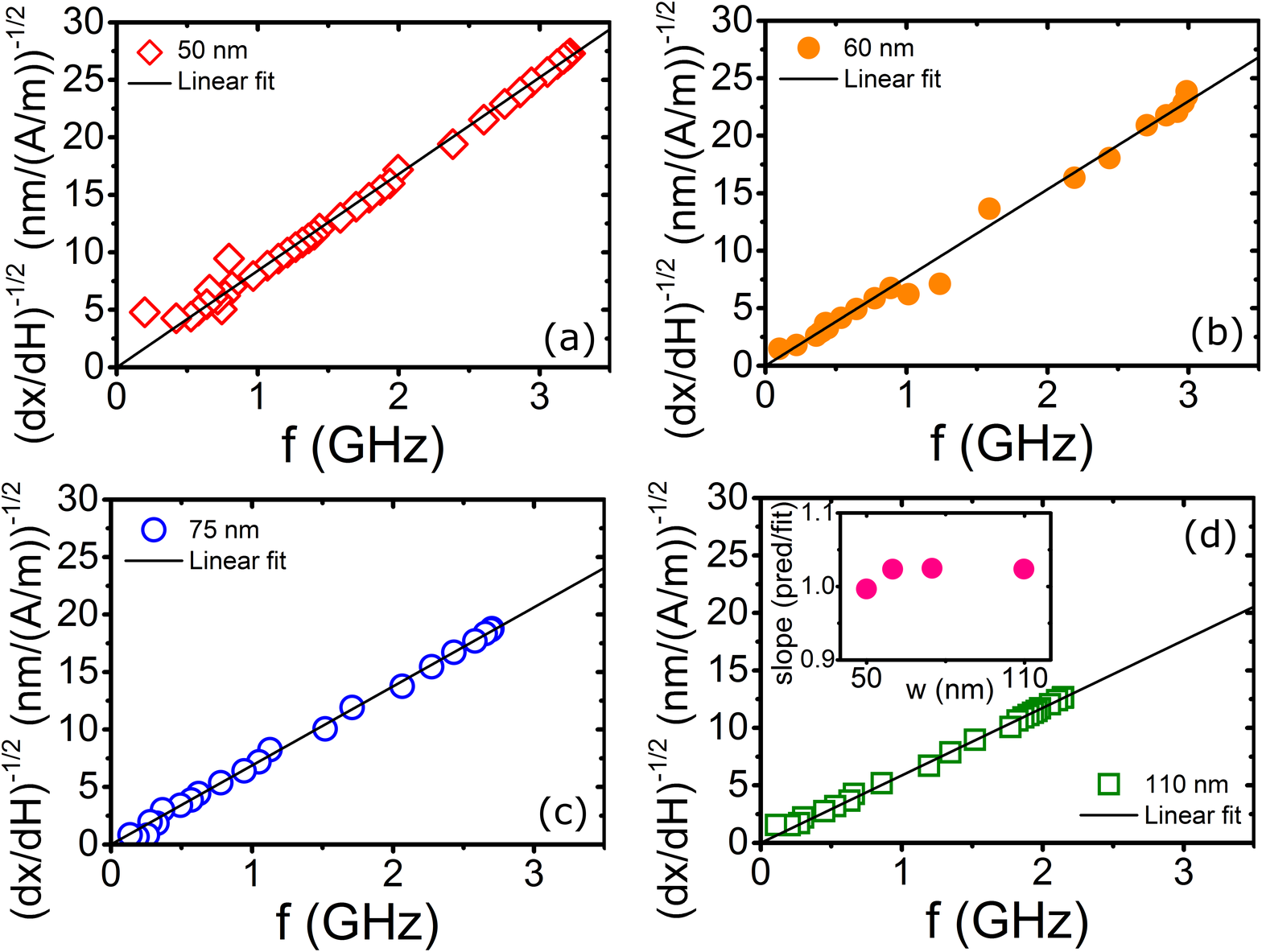}
	\caption{(Color online) Plot of $(dx/dH)^{-1/2}$, proportional to the square root of the local effective spring constant, versus $\ft$ for field-displaced TDWs in strip widths of (a) 50, (b) 60, (c) 75, and (d) 110 nm. $(dx/dH)^{-1}$ and $\ft$ data were taken, respectively, from Figs.~\ref{f:pinpot}(a) and \ref{f:depinning}. Solid lines are linear fits to the data assuming a zero x-axis intercept. The inset in (d) shows the ratio of the slope of the data in (a-d) predicted from the spring model to the measured slope.}
	\label{f:probing}
\end{figure*}

As $\hdepin$ is approached, $\ftw$ also drops in frequency [Fig.~\ref{f:depinning}(b)] which may, in part, be due to the wide part of the TDW being away from the upper notch (as per Fig.~\ref{f:shiftedwall}). This shifts the concentration of the dynamics at the upper edge of the strip away from the notch [Fig.~\ref{f:depinning}(c)]. We have already seen that strongly reducing the size of the upper notch for an undisplaced wall reduces $\ftw$ [Fig.~\ref{f:notch}(d)] and the case of the displaced wall is somewhat analogous as the upper part of the wall is now far from the notch (i.e.~we effectively have $\dn\rightarrow 0$).  Unlike $\ft$, $\ftw$ also remains finite near depinning. This is analogous to the finite $\ftw$ observed for $\dn=0$ in Fig.~\ref{f:notch}(b). The breathing mode again shows a very weak change in its frequency even  near depinning where the  spatial profile of the mode is strongly deformed [Fig.~\ref{f:depinning}(d)] with respect to the case of a non-displaced wall  [Fig.~\ref{fig:modes}(c)]. Once again this highlights the robustness of $\fb$ to geometrical (and in-plane-field-induced) changes.

In Fig.~\ref{f:depinning} an increased $\ft$ can be observed at small strip widths (a trend which has already been seen in Fig.~\ref{f:widths}(a)) and this is accompanied by an increased $\hdepin$. To understand this, we will take the previously used approach of modeling a parabolic, notch-induced TDW confining potential\cite{Lepadatu2010,Martinez2007a,Martinez2008}. This results in a spring like behavior of the DW with a  restoring force of $-k_N x$ where $k_N$ is the pinned TDW's spring constant and $x$ its displacement from the center of the strip. The equilibrium position of the TDW at a given $H$ is  determined by a balance between this restoring force and the effective force due to the applied magnetic field  \cite{Martinez2007a,Martinez2008}.   This force can be estimated from the $x$-derivative of the change in Zeeman energy for the displaced TDW: $2 \mu_0 w t M_S H$ where $t=5$ nm is the strip thickness and  $\mu_0=4 \pi \times 10^{-7}$ H/m. Note that we neglect the locally altered strip width at the notch.

In Fig.~\ref{f:pinpot}(a) we plot the equilibrium position for the domain wall versus $H$ for the data shown in Fig.~\ref{f:depinning}. The position of the field-deformed TDW (see, e.g., Fig.~\ref{f:shiftedwall}), $x$, was determined from the spatially averaged $x$-component of the magnetization along the strip \cite{Martinez2007a,Martinez2008}. For low field, there is good linearity between $x$ and $H$, indicative of a close-to-parabolic pinning potential. At larger fields however, there is a faster than linear growth in the $\xtdw$, the effect of which will be discussed further below.
From the data in the linear region (which has slope $d x /dH=g_{\mathrm{linear}}$), we can estimate a value for $k_N$:
\begin{equation}
k_N=(2 \mu_0 w t M_S)(x/H)^{-1}=(2 \mu_0 w t M_S)g_{\mathrm{linear}}^{-1}.
\label{eq:}
\end{equation}
$k_N$, plotted in Fig.~\ref{f:pinpot}(b) versus the strip width, reduces with increasing strip width. At small widths, this results in a stiffer domain wall where the  notch (which has a fixed size  here) makes a larger \textit{relative} intrusion into the strip.

We can now use the values of $k_N$ to estimate $\ft$ at $H=0$ and compare to the data in Fig.~\ref{f:widths} \cite{Martinez2007a,Martinez2008}:
\begin{equation}
\ft=\frac{1}{2\pi}\sqrt{\frac{k_N}{m_w}}.
\label{eq:translationalfreq}
\end{equation}
Here, $m_w$ is the mass (e.g.~\cite{Saitoh2004,Tatara2004,Rhensius2010}) of the TDW. The increased $k_N$ at smaller strip widths [Fig.~\ref{f:pinpot}(b)] generates an increased resonant frequency as per  Fig.~\ref{f:depinning}. Thus the results are qualitatively consistent with the trend suggested by Eq.~(\ref{eq:translationalfreq}) under the assumption of a $w$-independent mass. To obtain numerical values for $\ft$ however, we must estimate the mass for which we use the damping-free ($\alpha=0$) expression\cite{Martinez2007a,Martinez2008} (a similar expression is given by Kr\"uger\cite{KrugerThesis}):
\begin{equation}
m_w=\frac{2\mu_0 w t}{\gamma^2(N_z-N_y)\Delta} \; .
\end{equation}
$\gamma=2.210713 \times 10^{5}$ m/A.s and $\Delta=\Delta_T$ is the Thiele DW width\cite{Thiaville2007}, $\Delta=\Delta_T$, defined by $2/\Delta_T=1/(wt)\int_V (dm/dx)^2$  with $V$ the nanostrip volume. We note that  the field-induced deformation of the TDW leads to an increased $\Delta_T$ [Fig.~\ref{f:pinpot}(c)]. $N_y$ and $N_z$ are the demagnetizing factors for the TDW in the $y$ and $z$ directions.  To calculate these factors, we used expressions given by  Aharoni\cite{Aharoni1998}, treating the TDW as a uniformly magnetized slab with a length in the $y$ direction equal to the strip width, a height in the $z$ direction equal to the strip thickness and a width in the $x$ direction of  $\rho \Delta_T(H=0)$. $\rho$, a scaling factor, is the only free parameter since the strip width and strip thickness are fixed. It sets the width of the rectangular prism used for the demagnetizing field calculation as a fraction of the Thiele width.

As can be seen in Fig.~\ref{f:pinpot}(d), good agreement between the eigenmode simulation at $H=0$ and the spring model [Eq.~(2)] is found for the four studied thicknesses when using $\rho=\frac{1}{3}$.  This means that the slab used for  the demagnetizing factor calculation is $\sim 10$ nm wide in the $x$ direction, essentially covering a central narrow slice of the TDW structure where the magnetization is quasi-uniformly magnetized in the y-direction [Fig.~\ref{f:lpin}(a)] and thus close to our original model of a uniformly magnetized slab. Note  that the  magnetization undergoes an almost complete rotation from being aligned along $+x$ to $-x$ over a much larger distance  $\sim\pi\Delta_T$ [Fig.~\ref{f:lpin}(a)]. It is  however the central region of the TDW  appears to be the relevant part in this approach.  

The effective width of the pinning potential, $L_{\mathrm{pin}}$, defined here as the maximum displacement of the TDW measured before depinning (Fig.~\ref{f:pinpot}(a)) increases with $\Delta_T$  [read from Fig.~\ref{f:lpin}(b)]. Although the width the pinning potential increases with strip width by a factor of $\sim 2$ over the range of studied strip widths, the spring constant decreases by a factor of $\sim 4$ suggesting that it is $k_N$ (rather than $L_{\mathrm{pin}}$) which is dominant in determining the depinning field which, like  $\ft$, decreases with strip width.

Finally, we address the faster than linear growth in the TDW position versus $H$ [Fig.~\ref{f:depinning}(a)] which is a result of the pinning potential having a reduced steepness near its edge.  We can  show that Eq.~(\ref{eq:translationalfreq})  remains valid in describing $\ft$ at $x \ne 0$ (i.e.~even in the  non-parabolic part\cite{Moriya2008} of the potential) if we replace   $k_N$  by a local effective spring constant 
\begin{equation} 
k_{N,\mathrm{eff}}(x(H)) =\frac{2 \mu_0 w t M_S}{d x /dH}.
\end{equation}
In Fig.~\ref{f:probing} $\sqrt{(d\xtdw/dH)^{-1}}$ ($\propto \ft$ as per Eq.~(\ref{eq:translationalfreq})) has been plotted  versus the simulated values of $\ft$  for all studied strips. We have neglected any field-induced change in the TDW mass ($m_w=m_w(H=0)$) and have used a numerical derivative  of the data in Fig.~\ref{f:pinpot}(a) to determine $d\xtdw/dH$. We find a high degree of linearity over the full field range for all strip widths. This confirms the continued  validity of Eq.~(\ref{eq:translationalfreq}) and demonstrates   that the sharp drop-off in $\ft$ near $\hdepin$ (Fig.~\ref{f:depinning}) can be  linked with a change in the local gradient of the pinning potential at its edge, the latter determining  the resonant frequency of the displaced TDW in the small oscillation limit.  Note that from Eq.~(\ref{eq:translationalfreq}), we expect that the slope of the data in Fig.~\ref{f:probing} will be $2\pi\sqrt{m_w(H=0)/2\mu_0 w t M_s}$.  We have plotted the ratio of the predicted slope to the fitted slope in  the inset of Fig.~\ref{f:probing}(d) where we indeed find consistency to within 2.5\%.

\section{Conclusion}

We have numerically calculated eigenmodes of transverse domain walls (TDWs) which are pinned at triangular notches in  in-plane magnetized nanostrips. This enabled the study of translational, twisting and breathing resonances of TDWs and the effect that notch geometry and field-induced TDW displacements have on these modes. 

The twisting and translational modes both involve either local or global lateral translation of the wall structure within the notch-induced pinning potential. This leads to a clear sensitivity to changes in the intrusion depth of the notches especially to that of the notch at the narrow end of the TDW structure which has a  dominant role in laterally confining the TDW. The breathing mode, which is characterized by dynamics concentrated at the edge of the TDW (and thus away from the notches), was relatively insensitive to changes in the notch intrusion depth. For example, when varying the notch intrusion depth from 0 to 20 nm, the largest change in the mode's frequency was 3\% (observed for the narrowest strip width of 60 nm).  The sensitivity could be further reduced by using a thin or wide strip. 

These results may  be relevant when choosing which TDW mode to exploit in DW oscillators or when aiming to individually or simultaneously excite  (multiple) DWs pinned at different positions within a strip (e.g. in shift registers)\cite{Parkin2008,Kim2014,Matsushita2014}. This is because certain modes (i.e.~those with a translational nature) will be more sensitive to non-uniformity of  notch geometries and/or to the presence of small uncontrolled defects. 
Our results suggest that the breathing mode frequency will be the most robust to the introduction of small unwanted defects or non-uniformity in fabricated notch geometries, especially at larger strip widths. In contrast, having a translational or twisting mode frequency which is robust to small changes in the notch geometry appears to be reliant on having relatively large notches. 

For a fixed notch geometry, the frequencies of  all modes increased with decreasing strip width, making this an important parameter to control in devices. For the translational mode, the width dependence could be reproduced with a spring model for the notch-induced TDW confinement. Indeed, the eigenmode method (which does \textit{not} rely on the forced driving of the TDW's resonant dynamics over time) allowed us to determine the translational mode frequencies over a wide range of fields, including in the close vicinity of the static depinning field where the translational mode frequency drops sharply towards zero. At low fields and thus low TDW displacements, the notch-induced confining potential was found to be parabolic, enabling us to analytically reproduce the simulated  translational mode frequency at zero field. At large fields and large displacements however, the growth of the pinning potential's energy with displacement was found to be sub-parabolic. Here the spring model could however still be used to reproduce the translational mode frequencies as long as the local slope of the pinning potential was used to calculate the spring constant. 

Finally, we note that $\ft$ is finite only in the presence of confinement. In contrast, $\fb$ and $\ftw$ remain large and finite even without a notch or close to depinning, suggesting that they correspond to intrinsic localized TDW excitations which do albeit show some degree of sensitivity to the presence of notches.

\begin{acknowledgments}
This research was supported by the French ANR grant ESPERADO
(11-BS10-008), the Australian Research Council's Discovery Early Career Researcher Award funding scheme (DE120100155), an EPSRC Doctoral Training Centre grant (EP/G03690X/1) and the University of Western Australia's Research Collaboration Award and Early Career Researcher Fellowship Support programs. PJM also acknowledges support from the United States Air Force, Asian Office of Aerospace Research and Development (AOARD). The authors thank M.~Kostylev, I.S.~Maksymov, T.~Valet and G.~Albuquerque for useful discussions.   
\end{acknowledgments}

\appendix

\section{Comparison to time domain `ringdown' simulations}\label{ringdown}

For consistency, we confirmed that the frequencies of the translational and breathing modes obtained via a time domain ringdown method were consistent with those from the eigenmode method. An excitation field was applied in the $x$ direction to determine $\ft$ and in the $y$ direction to determine $\ftw$. Experimentally, this could be achieved using microwave frequency $x$ or $y$ oriented (real or effective) magnetic fields generated by striplines \cite{Bocklage2014},  Oersted fields due to in-plane current injection \cite{Uhlir2011} or tailorable spin torques in magnetoresistive devices \cite{Chanthbouala2011,Khvalkovskiy2009,Boone2010,Metaxas2013b,Lequeux2015}. Fourier analysis of the ringdown dynamics at a strip width of 80 nm demonstrated excitation of the translational and breathing modes at $\ft=2.6\pm0.1$ GHz and $\fb=6.4\pm0.1 $ GHz, in good agreement with the eigenmode results ($\ft=2.61$ GHz and $\fb=6.38$ GHz for $w=80$ nm as per Fig.~\ref{f:widths}).    Note that we were unable to efficiently excite the twisting mode using spatially uniform excitations along the $x$, $y$ and diagonal axes or a non-uniform field parallel to the $x$-axis everywhere with strength proportional to the y-position; i.e.~pointing in positive (negative) $x$-direction at positive (negative) $y$ as per Fig.~\ref{fig:modes}(a). 

\begin{figure}
	\includegraphics[width=6cm]{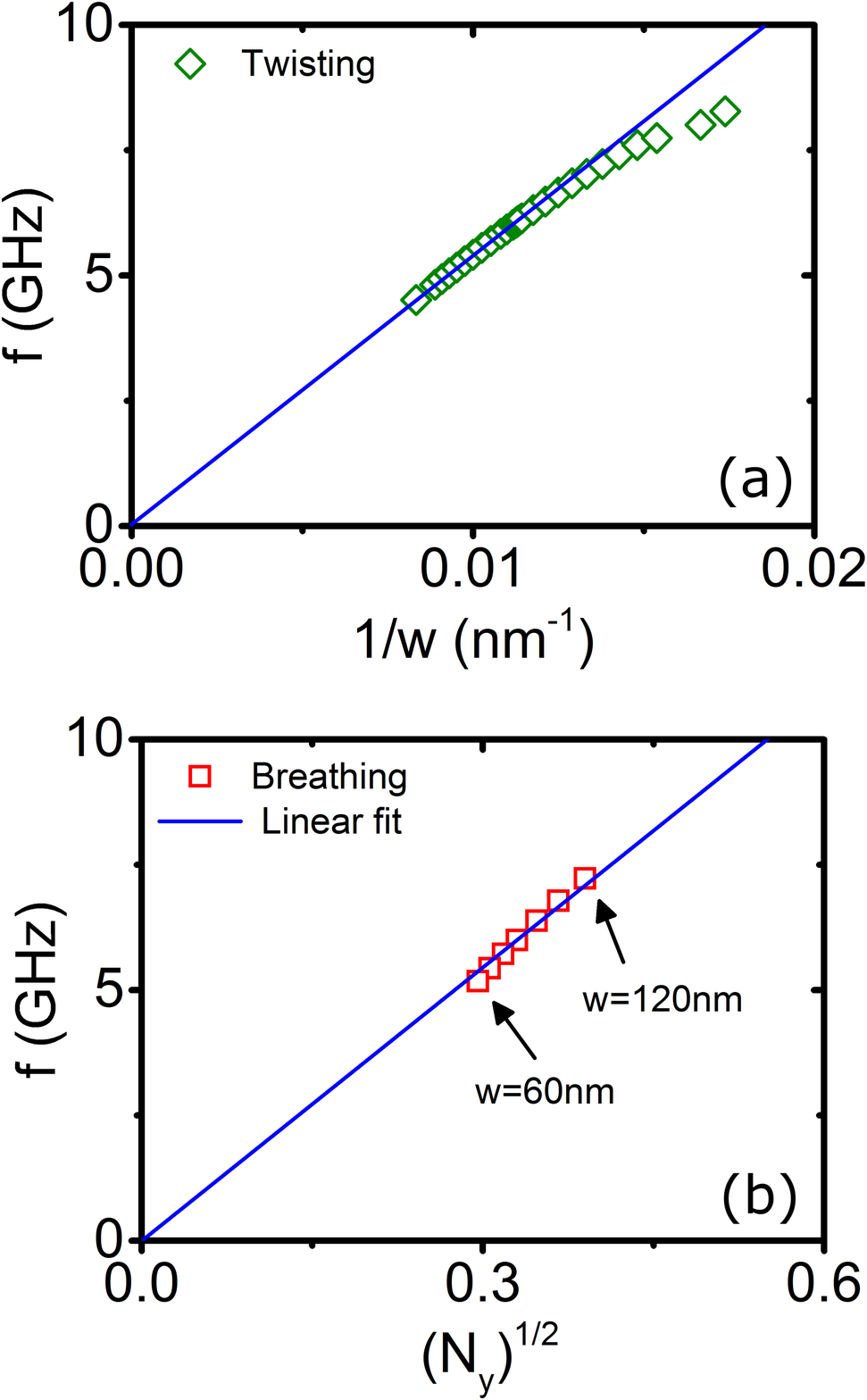}
	\caption{(Color online) (a) $\ftw$ versus the inverse strip width. (b) $\fb$ versus $\sqrt{N_y}$ (see text for $N_y$ calculation) for a number of strip widths. The linear fits have been obtained by constraining the x-axis intercept to zero. }
	\label{f:simpletest}
\end{figure}

\section{Extraction of pure modes from mixed modes}\label{mixing}

To demonstrate that each mixed mode  is a linear combination of the `pure' orthogonal twisting and breathing eigenmodes,
  we let $\v_1, \v_2$ be the `mixed mode' eigenvectors as returned by the
  solver (their complex entries encode the amplitude and relative
  phase of the magnetization oscillations at each mesh node). To show
  that these can be reduced to the `pure' modes we need to find
  complex scalars $a_1, a_2$ such that the linear combination $\v = a_1
  \v_1 + a_2 \v_2$ represents a breathing/twisting mode. The breathing
  mode is characterized by being fully symmetric about the $y$-axis,
  i.e.~the oscillations in the left and right half of the nanostrip
  are out of phase by $180^\circ$: $\v(x, y, z) = -\v(-x, y, z)$. The
  expression $\int \left| \v(x, y, z) + \v(-x, y, z) \right|$ thus measures
  the deviation from symmetry for an eigenmode $\v$ and we can find
  the `most symmetric' linear combination by minimizing this with
  respect to $a_1, a_2$. Since each eigenvector is only determined up
  to a scalar, we can assume that $a_1=1$ (or $a_2=1$), reducing the
  dimensionality of the optimization problem. The obtained linear
  combination is confirmed to be an eigenvector corresponding to a
  breathing mode. Similarly, the twisting mode can be recovered by
  using the condition $\v_{\mathrm{twist}}(x, y, z) =
  \v_{\mathrm{twist}}(-x, y, z)$.

\section{Modeling the twisting and breathing modes}\label{othermodes}

We detail here two simple qualitative models for the $\fb$ and$\ftw$ strip width dependencies seen in Fig.~\ref{f:widths}(a). 
The  general trend of decreasing $\ftw$ with $w$ for fixed notch geometry is qualitatively consistent with a string-like mode that is confined across the strip having a single node in the strip's center (i.e.~with wavelength $\sim 2w$ and thus a frequency $\propto \frac{1}{w}$). We plot $\ftw$~versus $\frac{1}{w}$ in Fig.~\ref{f:simpletest}(a) with reasonable linearity at larger widths.  
Liu and Gr\"utter have constructed a model for DW width resonances in magnetic films \cite{Liu1998} which predicts that $\fb$ will be proportional to $\sqrt{K_{eff}}$ where $K_{eff}$ is the effective anisotropy energy associated with the domain wall. For our static TDW (here in a confined geometry rather than a continuous layer), $K_{eff}$ comes from the TDW's demagnetizing energy and can be written as $\frac{1}{2}\mu_0M_S^2N_{y}$ (e.g.~\cite{Bryan2012}), giving $\fb\propto\sqrt{N_{y}}$. Indeed, this relation reproduces the observed $\fb$ trend relatively well over the entire strip width range, as calculated for a number of strip width values in Fig.~\ref{f:simpletest}. To determine $N_{y}$, we used the same slab approach as  used in Sec.~\ref{widthsection}.


\begin{thebibliography}{62}%
\makeatletter
\providecommand \@ifxundefined [1]{%
 \@ifx{#1\undefined}
}%
\providecommand \@ifnum [1]{%
 \ifnum #1\expandafter \@firstoftwo
 \else \expandafter \@secondoftwo
 \fi
}%
\providecommand \@ifx [1]{%
 \ifx #1\expandafter \@firstoftwo
 \else \expandafter \@secondoftwo
 \fi
}%
\providecommand \natexlab [1]{#1}%
\providecommand \enquote  [1]{``#1''}%
\providecommand \bibnamefont  [1]{#1}%
\providecommand \bibfnamefont [1]{#1}%
\providecommand \citenamefont [1]{#1}%
\providecommand \href@noop [0]{\@secondoftwo}%
\providecommand \href [0]{\begingroup \@sanitize@url \@href}%
\providecommand \@href[1]{\@@startlink{#1}\@@href}%
\providecommand \@@href[1]{\endgroup#1\@@endlink}%
\providecommand \@sanitize@url [0]{\catcode `\\12\catcode `\$12\catcode
  `\&12\catcode `\#12\catcode `\^12\catcode `\_12\catcode `\%12\relax}%
\providecommand \@@startlink[1]{}%
\providecommand \@@endlink[0]{}%
\providecommand \url  [0]{\begingroup\@sanitize@url \@url }%
\providecommand \@url [1]{\endgroup\@href {#1}{\urlprefix }}%
\providecommand \urlprefix  [0]{URL }%
\providecommand \Eprint [0]{\href }%
\providecommand \doibase [0]{http://dx.doi.org/}%
\providecommand \selectlanguage [0]{\@gobble}%
\providecommand \bibinfo  [0]{\@secondoftwo}%
\providecommand \bibfield  [0]{\@secondoftwo}%
\providecommand \translation [1]{[#1]}%
\providecommand \BibitemOpen [0]{}%
\providecommand \bibitemStop [0]{}%
\providecommand \bibitemNoStop [0]{.\EOS\space}%
\providecommand \EOS [0]{\spacefactor3000\relax}%
\providecommand \BibitemShut  [1]{\csname bibitem#1\endcsname}%
\let\auto@bib@innerbib\@empty
\bibitem [{\citenamefont {Parkin}\ \emph {et~al.}(2008)\citenamefont {Parkin},
  \citenamefont {Hayashi},\ and\ \citenamefont {Thomas}}]{Parkin2008}%
  \BibitemOpen
  \bibfield  {author} {\bibinfo {author} {\bibfnamefont {S.~S.~P.}\
  \bibnamefont {Parkin}}, \bibinfo {author} {\bibfnamefont {M.}~\bibnamefont
  {Hayashi}}, \ and\ \bibinfo {author} {\bibfnamefont {L.}~\bibnamefont
  {Thomas}},\ }\href@noop {} {\bibfield  {journal} {\bibinfo  {journal}
  {Science}\ }\textbf {\bibinfo {volume} {320}},\ \bibinfo {pages} {190}
  (\bibinfo {year} {2008})}\BibitemShut {NoStop}%
\bibitem [{\citenamefont {Wang}\ \emph {et~al.}(2009)\citenamefont {Wang},
  \citenamefont {Chen}, \citenamefont {Xi}, \citenamefont {Li},\ and\
  \citenamefont {Dimitrov}}]{Wang2009}%
  \BibitemOpen
  \bibfield  {author} {\bibinfo {author} {\bibfnamefont {X.}~\bibnamefont
  {Wang}}, \bibinfo {author} {\bibfnamefont {Y.}~\bibnamefont {Chen}}, \bibinfo
  {author} {\bibfnamefont {H.}~\bibnamefont {Xi}}, \bibinfo {author}
  {\bibfnamefont {H.}~\bibnamefont {Li}}, \ and\ \bibinfo {author}
  {\bibfnamefont {D.}~\bibnamefont {Dimitrov}},\ }\href@noop {} {\bibfield
  {journal} {\bibinfo  {journal} {IEEE Elec. Dev. Lett.}\ }\textbf {\bibinfo
  {volume} {30}},\ \bibinfo {pages} {294} (\bibinfo {year} {2009})}\BibitemShut
  {NoStop}%
\bibitem [{\citenamefont {Locatelli}\ \emph {et~al.}(2013)\citenamefont
  {Locatelli}, \citenamefont {Cros},\ and\ \citenamefont
  {Grollier}}]{Locatelli2013}%
  \BibitemOpen
  \bibfield  {author} {\bibinfo {author} {\bibfnamefont {N.}~\bibnamefont
  {Locatelli}}, \bibinfo {author} {\bibfnamefont {V.}~\bibnamefont {Cros}}, \
  and\ \bibinfo {author} {\bibfnamefont {J.}~\bibnamefont {Grollier}},\ }\href
  {\doibase 10.1038/nmat3823} {\bibfield  {journal} {\bibinfo  {journal} {Nat.
  Mater.}\ }\textbf {\bibinfo {volume} {13}},\ \bibinfo {pages} {11} (\bibinfo
  {year} {2013})}\BibitemShut {NoStop}%
\bibitem [{\citenamefont {Donolato}\ \emph {et~al.}(2010)\citenamefont
  {Donolato}, \citenamefont {Vavassori}, \citenamefont {Gobbi}, \citenamefont
  {Deryabina}, \citenamefont {Hansen}, \citenamefont {Metlushko}, \citenamefont
  {Ilic}, \citenamefont {Cantoni}, \citenamefont {Petti}, \citenamefont
  {Brivio},\ and\ \citenamefont {et~al.}}]{Donolato2010}%
  \BibitemOpen
  \bibfield  {author} {\bibinfo {author} {\bibfnamefont {M.}~\bibnamefont
  {Donolato}}, \bibinfo {author} {\bibfnamefont {P.}~\bibnamefont {Vavassori}},
  \bibinfo {author} {\bibfnamefont {M.}~\bibnamefont {Gobbi}}, \bibinfo
  {author} {\bibfnamefont {M.}~\bibnamefont {Deryabina}}, \bibinfo {author}
  {\bibfnamefont {M.~F.}\ \bibnamefont {Hansen}}, \bibinfo {author}
  {\bibfnamefont {V.}~\bibnamefont {Metlushko}}, \bibinfo {author}
  {\bibfnamefont {B.}~\bibnamefont {Ilic}}, \bibinfo {author} {\bibfnamefont
  {M.}~\bibnamefont {Cantoni}}, \bibinfo {author} {\bibfnamefont
  {D.}~\bibnamefont {Petti}}, \bibinfo {author} {\bibfnamefont
  {S.}~\bibnamefont {Brivio}}, \ and\ \bibinfo {author} {\bibnamefont
  {et~al.}},\ }\href {\doibase 10.1002/adma.201000146} {\bibfield  {journal}
  {\bibinfo  {journal} {Adv. Mater.}\ }\textbf {\bibinfo {volume} {22}},\
  \bibinfo {pages} {2706} (\bibinfo {year} {2010})}\BibitemShut {NoStop}%
\bibitem [{\citenamefont {Rapoport}\ \emph {et~al.}(2012)\citenamefont
  {Rapoport}, \citenamefont {Montana},\ and\ \citenamefont
  {Beach}}]{Rapoport2012a}%
  \BibitemOpen
  \bibfield  {author} {\bibinfo {author} {\bibfnamefont {E.}~\bibnamefont
  {Rapoport}}, \bibinfo {author} {\bibfnamefont {D.}~\bibnamefont {Montana}}, \
  and\ \bibinfo {author} {\bibfnamefont {G.~S.~D.}\ \bibnamefont {Beach}},\
  }\href {\doibase 10.1039/c2lc40715a} {\bibfield  {journal} {\bibinfo
  {journal} {Lab Chip}\ }\textbf {\bibinfo {volume} {12}},\ \bibinfo {pages}
  {4433} (\bibinfo {year} {2012})}\BibitemShut {NoStop}%
\bibitem [{\citenamefont {Saitoh}\ \emph {et~al.}(2004)\citenamefont {Saitoh},
  \citenamefont {Miyajima}, \citenamefont {Yamaoka},\ and\ \citenamefont
  {Tatara}}]{Saitoh2004}%
  \BibitemOpen
  \bibfield  {author} {\bibinfo {author} {\bibfnamefont {E.}~\bibnamefont
  {Saitoh}}, \bibinfo {author} {\bibfnamefont {H.}~\bibnamefont {Miyajima}},
  \bibinfo {author} {\bibfnamefont {T.}~\bibnamefont {Yamaoka}}, \ and\
  \bibinfo {author} {\bibfnamefont {G.}~\bibnamefont {Tatara}},\ }\href@noop {}
  {\bibfield  {journal} {\bibinfo  {journal} {Nature}\ }\textbf {\bibinfo
  {volume} {432}},\ \bibinfo {pages} {203} (\bibinfo {year}
  {2004})}\BibitemShut {NoStop}%
\bibitem [{\citenamefont {Winter}(1961)}]{Winter1961}%
  \BibitemOpen
  \bibfield  {author} {\bibinfo {author} {\bibfnamefont {J.}~\bibnamefont
  {Winter}},\ }\href@noop {} {\bibfield  {journal} {\bibinfo  {journal}
  {Physical Review}\ }\textbf {\bibinfo {volume} {124}},\ \bibinfo {pages}
  {452} (\bibinfo {year} {1961})}\BibitemShut {NoStop}%
\bibitem [{\citenamefont {Rebei}\ and\ \citenamefont
  {Mryasov}(2006)}]{Rebei2006}%
  \BibitemOpen
  \bibfield  {author} {\bibinfo {author} {\bibfnamefont {A.}~\bibnamefont
  {Rebei}}\ and\ \bibinfo {author} {\bibfnamefont {O.}~\bibnamefont
  {Mryasov}},\ }\href@noop {} {\bibfield  {journal} {\bibinfo  {journal} {Phys.
  Rev. B}\ }\textbf {\bibinfo {volume} {74}},\ \bibinfo {pages} {014412}
  (\bibinfo {year} {2006})}\BibitemShut {NoStop}%
\bibitem [{\citenamefont {Bedau}\ \emph {et~al.}(2007)\citenamefont {Bedau},
  \citenamefont {Kla{\"u}i}, \citenamefont {Krzyk}, \citenamefont
  {R{\"u}diger}, \citenamefont {Faini},\ and\ \citenamefont
  {Vila}}]{Bedau2007}%
  \BibitemOpen
  \bibfield  {author} {\bibinfo {author} {\bibfnamefont {D.}~\bibnamefont
  {Bedau}}, \bibinfo {author} {\bibfnamefont {M.}~\bibnamefont {Kla{\"u}i}},
  \bibinfo {author} {\bibfnamefont {S.}~\bibnamefont {Krzyk}}, \bibinfo
  {author} {\bibfnamefont {U.}~\bibnamefont {R{\"u}diger}}, \bibinfo {author}
  {\bibfnamefont {G.}~\bibnamefont {Faini}}, \ and\ \bibinfo {author}
  {\bibfnamefont {L.}~\bibnamefont {Vila}},\ }\href@noop {} {\bibfield
  {journal} {\bibinfo  {journal} {Phys. Rev. Lett.}\ }\textbf {\bibinfo
  {volume} {99}},\ \bibinfo {pages} {146601} (\bibinfo {year}
  {2007})}\BibitemShut {NoStop}%
\bibitem [{\citenamefont {Sandweg}\ \emph {et~al.}(2008)\citenamefont
  {Sandweg}, \citenamefont {Hermsdoerfer}, \citenamefont {Schultheiss},
  \citenamefont {Sch\"afer}, \citenamefont {Leven},\ and\ \citenamefont
  {Hillebrands}}]{Sandweg2008}%
  \BibitemOpen
  \bibfield  {author} {\bibinfo {author} {\bibfnamefont {C.~W.}\ \bibnamefont
  {Sandweg}}, \bibinfo {author} {\bibfnamefont {S.~J.}\ \bibnamefont
  {Hermsdoerfer}}, \bibinfo {author} {\bibfnamefont {H.}~\bibnamefont
  {Schultheiss}}, \bibinfo {author} {\bibfnamefont {R.}~\bibnamefont
  {Sch\"afer}}, \bibinfo {author} {\bibfnamefont {B.}~\bibnamefont {Leven}}, \
  and\ \bibinfo {author} {\bibfnamefont {B.}~\bibnamefont {Hillebrands}},\
  }\href@noop {} {\bibfield  {journal} {\bibinfo  {journal} {J. Phys. D: Appl.
  Phys.}\ }\textbf {\bibinfo {volume} {41}},\ \bibinfo {pages} {164008}
  (\bibinfo {year} {2008})}\BibitemShut {NoStop}%
\bibitem [{\citenamefont {Roy}\ \emph {et~al.}(2010)\citenamefont {Roy},
  \citenamefont {Trypiniotis},\ and\ \citenamefont {Barnes}}]{Roy2010}%
  \BibitemOpen
  \bibfield  {author} {\bibinfo {author} {\bibfnamefont {P.~E.}\ \bibnamefont
  {Roy}}, \bibinfo {author} {\bibfnamefont {T.}~\bibnamefont {Trypiniotis}}, \
  and\ \bibinfo {author} {\bibfnamefont {C.~H.~W.}\ \bibnamefont {Barnes}},\
  }\href {\doibase 10.1103/PhysRevB.82.134411} {\bibfield  {journal} {\bibinfo
  {journal} {Phys. Rev. B}\ }\textbf {\bibinfo {volume} {82}},\ \bibinfo
  {pages} {134411} (\bibinfo {year} {2010})}\BibitemShut {NoStop}%
\bibitem [{\citenamefont {Sangiao}\ and\ \citenamefont
  {Viret}(2014)}]{Sangiao2014}%
  \BibitemOpen
  \bibfield  {author} {\bibinfo {author} {\bibfnamefont {S.}~\bibnamefont
  {Sangiao}}\ and\ \bibinfo {author} {\bibfnamefont {M.}~\bibnamefont
  {Viret}},\ }\href@noop {} {\bibfield  {journal} {\bibinfo  {journal} {Phys.
  Rev. B}\ }\textbf {\bibinfo {volume} {89}},\ \bibinfo {pages} {104412}
  (\bibinfo {year} {2014})}\BibitemShut {NoStop}%
\bibitem [{\citenamefont {Lequeux}\ \emph {et~al.}(2015)\citenamefont
  {Lequeux}, \citenamefont {Sampaio}, \citenamefont {Bortolotti}, \citenamefont
  {Devolder}, \citenamefont {Matsumoto}, \citenamefont {Yakushiji},
  \citenamefont {Kubota}, \citenamefont {Fukushima}, \citenamefont {Yuasa},
  \citenamefont {Nishimura}, \citenamefont {Nagamine}, \citenamefont
  {Tsunekawa}, \citenamefont {Cros},\ and\ \citenamefont
  {Grollier}}]{Lequeux2015}%
  \BibitemOpen
  \bibfield  {author} {\bibinfo {author} {\bibfnamefont {S.}~\bibnamefont
  {Lequeux}}, \bibinfo {author} {\bibfnamefont {J.}~\bibnamefont {Sampaio}},
  \bibinfo {author} {\bibfnamefont {P.}~\bibnamefont {Bortolotti}}, \bibinfo
  {author} {\bibfnamefont {T.}~\bibnamefont {Devolder}}, \bibinfo {author}
  {\bibfnamefont {R.}~\bibnamefont {Matsumoto}}, \bibinfo {author}
  {\bibfnamefont {K.}~\bibnamefont {Yakushiji}}, \bibinfo {author}
  {\bibfnamefont {H.}~\bibnamefont {Kubota}}, \bibinfo {author} {\bibfnamefont
  {A.}~\bibnamefont {Fukushima}}, \bibinfo {author} {\bibfnamefont
  {S.}~\bibnamefont {Yuasa}}, \bibinfo {author} {\bibfnamefont
  {K.}~\bibnamefont {Nishimura}}, \bibinfo {author} {\bibfnamefont
  {Y.}~\bibnamefont {Nagamine}}, \bibinfo {author} {\bibfnamefont
  {K.}~\bibnamefont {Tsunekawa}}, \bibinfo {author} {\bibfnamefont
  {V.}~\bibnamefont {Cros}}, \ and\ \bibinfo {author} {\bibfnamefont
  {J.}~\bibnamefont {Grollier}},\ }\href@noop {} {\bibfield  {journal}
  {\bibinfo  {journal} {Pre-print}\ }\textbf {\bibinfo {volume}
  {arxiv:1508.04043}} (\bibinfo {year} {2015})}\BibitemShut {NoStop}%
\bibitem [{\citenamefont {Lepadatu}\ \emph {et~al.}(2010)\citenamefont
  {Lepadatu}, \citenamefont {Wesseley}, \citenamefont {Vanhaverbeke},
  \citenamefont {Allenspach}, \citenamefont {Potenza}, \citenamefont
  {Marchetto}, \citenamefont {Charlton}, \citenamefont {Langridge},
  \citenamefont {Dhesi},\ and\ \citenamefont {Marrows}}]{Lepadatu2010}%
  \BibitemOpen
  \bibfield  {author} {\bibinfo {author} {\bibfnamefont {S.}~\bibnamefont
  {Lepadatu}}, \bibinfo {author} {\bibfnamefont {O.}~\bibnamefont {Wesseley}},
  \bibinfo {author} {\bibfnamefont {A.}~\bibnamefont {Vanhaverbeke}}, \bibinfo
  {author} {\bibfnamefont {R.}~\bibnamefont {Allenspach}}, \bibinfo {author}
  {\bibfnamefont {A.}~\bibnamefont {Potenza}}, \bibinfo {author} {\bibfnamefont
  {H.}~\bibnamefont {Marchetto}}, \bibinfo {author} {\bibfnamefont {T.~R.}\
  \bibnamefont {Charlton}}, \bibinfo {author} {\bibfnamefont {S.}~\bibnamefont
  {Langridge}}, \bibinfo {author} {\bibfnamefont {S.~S.}\ \bibnamefont
  {Dhesi}}, \ and\ \bibinfo {author} {\bibfnamefont {C.~H.}\ \bibnamefont
  {Marrows}},\ }\href@noop {} {\bibfield  {journal} {\bibinfo  {journal} {Phys.
  Rev. B}\ }\textbf {\bibinfo {volume} {81}},\ \bibinfo {pages} {060402(R)}
  (\bibinfo {year} {2010})}\BibitemShut {NoStop}%
\bibitem [{\citenamefont {Bayer}\ \emph {et~al.}(2005)\citenamefont {Bayer},
  \citenamefont {Schultheiss}, \citenamefont {Hillebrands},\ and\ \citenamefont
  {Stamps}}]{Bayer2005}%
  \BibitemOpen
  \bibfield  {author} {\bibinfo {author} {\bibfnamefont {C.}~\bibnamefont
  {Bayer}}, \bibinfo {author} {\bibfnamefont {H.}~\bibnamefont {Schultheiss}},
  \bibinfo {author} {\bibfnamefont {B.}~\bibnamefont {Hillebrands}}, \ and\
  \bibinfo {author} {\bibfnamefont {R.~L.}\ \bibnamefont {Stamps}},\
  }\href@noop {} {\bibfield  {journal} {\bibinfo  {journal} {IEEE Trans. Mag.}\
  }\textbf {\bibinfo {volume} {41}},\ \bibinfo {pages} {3094} (\bibinfo {year}
  {2005})}\BibitemShut {NoStop}%
\bibitem [{\citenamefont {Hermsdoerfer}\ \emph {et~al.}(2009)\citenamefont
  {Hermsdoerfer}, \citenamefont {Schultheiss}, \citenamefont {Rausch},
  \citenamefont {Schaￌﾈfer}, \citenamefont {Leven}, \citenamefont {Kim},\
  and\ \citenamefont {Hillebrands}}]{Hermsdoerfer2009}%
  \BibitemOpen
  \bibfield  {author} {\bibinfo {author} {\bibfnamefont {S.~J.}\ \bibnamefont
  {Hermsdoerfer}}, \bibinfo {author} {\bibfnamefont {H.}~\bibnamefont
  {Schultheiss}}, \bibinfo {author} {\bibfnamefont {C.}~\bibnamefont {Rausch}},
  \bibinfo {author} {\bibfnamefont {S.}~\bibnamefont {Schaￌﾈfer}}, \bibinfo
  {author} {\bibfnamefont {B.}~\bibnamefont {Leven}}, \bibinfo {author}
  {\bibfnamefont {S.-K.}\ \bibnamefont {Kim}}, \ and\ \bibinfo {author}
  {\bibfnamefont {B.}~\bibnamefont {Hillebrands}},\ }\href {\doibase
  10.1063/1.3143225} {\bibfield  {journal} {\bibinfo  {journal} {Appl. Phys.
  Lett.}\ }\textbf {\bibinfo {volume} {94}},\ \bibinfo {pages} {223510}
  (\bibinfo {year} {2009})}\BibitemShut {NoStop}%
\bibitem [{\citenamefont {Le~Maho}\ \emph {et~al.}(2009)\citenamefont
  {Le~Maho}, \citenamefont {Kim},\ and\ \citenamefont {Tatara}}]{LeMaho2009}%
  \BibitemOpen
  \bibfield  {author} {\bibinfo {author} {\bibfnamefont {Y.}~\bibnamefont
  {Le~Maho}}, \bibinfo {author} {\bibfnamefont {J.-V.}\ \bibnamefont {Kim}}, \
  and\ \bibinfo {author} {\bibfnamefont {G.}~\bibnamefont {Tatara}},\ }\href
  {\doibase 10.1103/PhysRevB.79.174404} {\bibfield  {journal} {\bibinfo
  {journal} {Phys. Rev. B}\ }\textbf {\bibinfo {volume} {79}},\ \bibinfo
  {pages} {174404} (\bibinfo {year} {2009})}\BibitemShut {NoStop}%
\bibitem [{\citenamefont {Han}\ \emph {et~al.}(2009)\citenamefont {Han},
  \citenamefont {Kim}, \citenamefont {Lee}, \citenamefont {Hermsdoerfer},
  \citenamefont {Schultheiss}, \citenamefont {Leven},\ and\ \citenamefont
  {Hillebrands}}]{Han2009}%
  \BibitemOpen
  \bibfield  {author} {\bibinfo {author} {\bibfnamefont {D.~S.}\ \bibnamefont
  {Han}}, \bibinfo {author} {\bibfnamefont {S.~K.}\ \bibnamefont {Kim}},
  \bibinfo {author} {\bibfnamefont {J.~Y.}\ \bibnamefont {Lee}}, \bibinfo
  {author} {\bibfnamefont {S.~J.}\ \bibnamefont {Hermsdoerfer}}, \bibinfo
  {author} {\bibfnamefont {H.}~\bibnamefont {Schultheiss}}, \bibinfo {author}
  {\bibfnamefont {B.}~\bibnamefont {Leven}}, \ and\ \bibinfo {author}
  {\bibfnamefont {B.}~\bibnamefont {Hillebrands}},\ }\href@noop {} {\bibfield
  {journal} {\bibinfo  {journal} {Appl. Phys. Lett}\ }\textbf {\bibinfo
  {volume} {94}},\ \bibinfo {pages} {112502} (\bibinfo {year}
  {2009})}\BibitemShut {NoStop}%
\bibitem [{\citenamefont {Jamali}\ \emph {et~al.}(2010)\citenamefont {Jamali},
  \citenamefont {Yang},\ and\ \citenamefont {Lee}}]{Jamali2010}%
  \BibitemOpen
  \bibfield  {author} {\bibinfo {author} {\bibfnamefont {M.}~\bibnamefont
  {Jamali}}, \bibinfo {author} {\bibfnamefont {H.}~\bibnamefont {Yang}}, \ and\
  \bibinfo {author} {\bibfnamefont {K.~J.}\ \bibnamefont {Lee}},\ }\href@noop
  {} {\bibfield  {journal} {\bibinfo  {journal} {Appl. Phys. Lett.}\ }\textbf
  {\bibinfo {volume} {96}},\ \bibinfo {pages} {242501} (\bibinfo {year}
  {2010})}\BibitemShut {NoStop}%
\bibitem [{\citenamefont {Janutka}(2013)}]{Janutka2013}%
  \BibitemOpen
  \bibfield  {author} {\bibinfo {author} {\bibfnamefont {A.}~\bibnamefont
  {Janutka}},\ }\href@noop {} {\bibfield  {journal} {\bibinfo  {journal} {IEEE
  Mag. Lett.}\ }\textbf {\bibinfo {volume} {4}},\ \bibinfo {pages} {4000104}
  (\bibinfo {year} {2013})}\BibitemShut {NoStop}%
\bibitem [{\citenamefont {Wang}\ \emph {et~al.}(2014)\citenamefont {Wang},
  \citenamefont {Guo}, \citenamefont {Nie}, \citenamefont {Wang}, \citenamefont
  {Zeng}, \citenamefont {Li},\ and\ \citenamefont {Tang}}]{Wang2014}%
  \BibitemOpen
  \bibfield  {author} {\bibinfo {author} {\bibfnamefont {X.~G.}\ \bibnamefont
  {Wang}}, \bibinfo {author} {\bibfnamefont {G.~H.}\ \bibnamefont {Guo}},
  \bibinfo {author} {\bibfnamefont {Y.~Z.}\ \bibnamefont {Nie}}, \bibinfo
  {author} {\bibfnamefont {D.~W.}\ \bibnamefont {Wang}}, \bibinfo {author}
  {\bibfnamefont {Z.~M.}\ \bibnamefont {Zeng}}, \bibinfo {author}
  {\bibfnamefont {Z.~X.}\ \bibnamefont {Li}}, \ and\ \bibinfo {author}
  {\bibfnamefont {W.}~\bibnamefont {Tang}},\ }\href@noop {} {\bibfield
  {journal} {\bibinfo  {journal} {Phys. Rev. B}\ }\textbf {\bibinfo {volume}
  {89}},\ \bibinfo {pages} {144418} (\bibinfo {year} {2014})}\BibitemShut
  {NoStop}%
\bibitem [{\citenamefont {Thomas}\ \emph {et~al.}(2007)\citenamefont {Thomas},
  \citenamefont {Hayashi}, \citenamefont {Jiang}, \citenamefont {Moriya},
  \citenamefont {Rettner},\ and\ \citenamefont {Parkin}}]{Thomas2007}%
  \BibitemOpen
  \bibfield  {author} {\bibinfo {author} {\bibfnamefont {L.}~\bibnamefont
  {Thomas}}, \bibinfo {author} {\bibfnamefont {M.}~\bibnamefont {Hayashi}},
  \bibinfo {author} {\bibfnamefont {X.}~\bibnamefont {Jiang}}, \bibinfo
  {author} {\bibfnamefont {R.}~\bibnamefont {Moriya}}, \bibinfo {author}
  {\bibfnamefont {C.}~\bibnamefont {Rettner}}, \ and\ \bibinfo {author}
  {\bibfnamefont {S.}~\bibnamefont {Parkin}},\ }\href@noop {} {\bibfield
  {journal} {\bibinfo  {journal} {Science}\ }\textbf {\bibinfo {volume}
  {315}},\ \bibinfo {pages} {1553} (\bibinfo {year} {2007})}\BibitemShut
  {NoStop}%
\bibitem [{\citenamefont {Nozaki}\ \emph {et~al.}(2007)\citenamefont {Nozaki},
  \citenamefont {Maekawa}, \citenamefont {Mizuguchi}, \citenamefont
  {Shiraishi}, \citenamefont {andY Suzuki}, \citenamefont {Maehara},
  \citenamefont {Kasai},\ and\ \citenamefont {Ono}}]{Nozaki2007}%
  \BibitemOpen
  \bibfield  {author} {\bibinfo {author} {\bibfnamefont {T.}~\bibnamefont
  {Nozaki}}, \bibinfo {author} {\bibfnamefont {H.}~\bibnamefont {Maekawa}},
  \bibinfo {author} {\bibfnamefont {M.}~\bibnamefont {Mizuguchi}}, \bibinfo
  {author} {\bibfnamefont {M.}~\bibnamefont {Shiraishi}}, \bibinfo {author}
  {\bibfnamefont {T.~S.}\ \bibnamefont {andY Suzuki}}, \bibinfo {author}
  {\bibfnamefont {H.}~\bibnamefont {Maehara}}, \bibinfo {author} {\bibfnamefont
  {S.}~\bibnamefont {Kasai}}, \ and\ \bibinfo {author} {\bibfnamefont
  {T.}~\bibnamefont {Ono}},\ }\href@noop {} {\bibfield  {journal} {\bibinfo
  {journal} {Appl. Phys. Lett.}\ }\textbf {\bibinfo {volume} {91}},\ \bibinfo
  {pages} {082502} (\bibinfo {year} {2007})}\BibitemShut {NoStop}%
\bibitem [{\citenamefont {Martinez}\ \emph {et~al.}(2008)\citenamefont
  {Martinez}, \citenamefont {{Lopez-Dias}}, \citenamefont {Alejos},\ and\
  \citenamefont {Torres}}]{Martinez2008}%
  \BibitemOpen
  \bibfield  {author} {\bibinfo {author} {\bibfnamefont {E.}~\bibnamefont
  {Martinez}}, \bibinfo {author} {\bibfnamefont {L.}~\bibnamefont
  {{Lopez-Dias}}}, \bibinfo {author} {\bibfnamefont {O.}~\bibnamefont
  {Alejos}}, \ and\ \bibinfo {author} {\bibfnamefont {L.}~\bibnamefont
  {Torres}},\ }\href@noop {} {\bibfield  {journal} {\bibinfo  {journal} {Phys.
  Rev. B}\ }\textbf {\bibinfo {volume} {77}},\ \bibinfo {pages} {144417}
  (\bibinfo {year} {2008})}\BibitemShut {NoStop}%
\bibitem [{\citenamefont {Metaxas}\ \emph {et~al.}(2010)\citenamefont
  {Metaxas}, \citenamefont {Anane}, \citenamefont {Cros}, \citenamefont
  {Grollier}, \citenamefont {Deranlot}, \citenamefont {Lema\^{\i}tre},
  \citenamefont {Xavier}, \citenamefont {Ulysse}, \citenamefont {Faini},
  \citenamefont {Petroff},\ and\ \citenamefont {Fert}}]{Metaxas2010apl}%
  \BibitemOpen
  \bibfield  {author} {\bibinfo {author} {\bibfnamefont {P.~J.}\ \bibnamefont
  {Metaxas}}, \bibinfo {author} {\bibfnamefont {A.}~\bibnamefont {Anane}},
  \bibinfo {author} {\bibfnamefont {V.}~\bibnamefont {Cros}}, \bibinfo {author}
  {\bibfnamefont {J.}~\bibnamefont {Grollier}}, \bibinfo {author}
  {\bibfnamefont {C.}~\bibnamefont {Deranlot}}, \bibinfo {author}
  {\bibfnamefont {A.}~\bibnamefont {Lema\^{\i}tre}}, \bibinfo {author}
  {\bibfnamefont {S.}~\bibnamefont {Xavier}}, \bibinfo {author} {\bibfnamefont
  {C.}~\bibnamefont {Ulysse}}, \bibinfo {author} {\bibfnamefont
  {G.}~\bibnamefont {Faini}}, \bibinfo {author} {\bibfnamefont
  {F.}~\bibnamefont {Petroff}}, \ and\ \bibinfo {author} {\bibfnamefont
  {A.}~\bibnamefont {Fert}},\ }\href@noop {} {\bibfield  {journal} {\bibinfo
  {journal} {Appl. Phys. Lett.}\ }\textbf {\bibinfo {volume} {97}},\ \bibinfo
  {pages} {182506} (\bibinfo {year} {2010})}\BibitemShut {NoStop}%
\bibitem [{\citenamefont {Petit}\ \emph {et~al.}(2008)\citenamefont {Petit},
  \citenamefont {Jausovec}, \citenamefont {Read},\ and\ \citenamefont
  {Cowburn}}]{Petit2008}%
  \BibitemOpen
  \bibfield  {author} {\bibinfo {author} {\bibfnamefont {D.}~\bibnamefont
  {Petit}}, \bibinfo {author} {\bibfnamefont {A.~V.}\ \bibnamefont {Jausovec}},
  \bibinfo {author} {\bibfnamefont {D.}~\bibnamefont {Read}}, \ and\ \bibinfo
  {author} {\bibfnamefont {R.~P.}\ \bibnamefont {Cowburn}},\ }\href@noop {}
  {\bibfield  {journal} {\bibinfo  {journal} {J. Appl. Phys.}\ }\textbf
  {\bibinfo {volume} {103}},\ \bibinfo {pages} {114307} (\bibinfo {year}
  {2008})}\BibitemShut {NoStop}%
\bibitem [{\citenamefont {Bogart}\ \emph {et~al.}(2009)\citenamefont {Bogart},
  \citenamefont {Atkinson}, \citenamefont {{O'Shea}}, \citenamefont
  {McGrouther},\ and\ \citenamefont {McVitie}}]{Bogart2009}%
  \BibitemOpen
  \bibfield  {author} {\bibinfo {author} {\bibfnamefont {L.K.}~\bibnamefont
  {Bogart}}, \bibinfo {author} {\bibfnamefont {D.}~\bibnamefont {Atkinson}},
  \bibinfo {author} {\bibfnamefont {K.}~\bibnamefont {{O'Shea}}}, \bibinfo
  {author} {\bibfnamefont {D.}~\bibnamefont {McGrouther}}, \ and\ \bibinfo
  {author} {\bibfnamefont {S.}~\bibnamefont {McVitie}},\ }\href@noop {}
  {\bibfield  {journal} {\bibinfo  {journal} {Phys. Rev. B}\ }\textbf {\bibinfo
  {volume} {79}},\ \bibinfo {pages} {054414} (\bibinfo {year}
  {2009})}\BibitemShut {NoStop}%
\bibitem [{\citenamefont {Kunz}\ and\ \citenamefont {Priem}(2010)}]{Kunz2010}%
  \BibitemOpen
  \bibfield  {author} {\bibinfo {author} {\bibfnamefont {A.}~\bibnamefont
  {Kunz}}\ and\ \bibinfo {author} {\bibfnamefont {J.~D.}\ \bibnamefont
  {Priem}},\ }\href@noop {} {\bibfield  {journal} {\bibinfo  {journal} {IEEE
  Trans. Mag.}\ }\textbf {\bibinfo {volume} {46}},\ \bibinfo {pages} {1559}
  (\bibinfo {year} {2010})}\BibitemShut {NoStop}%
\bibitem [{\citenamefont {Currivan}\ \emph {et~al.}(2014)\citenamefont
  {Currivan}, \citenamefont {Siddiqui}, \citenamefont {Ahn}, \citenamefont
  {Tryputen}, \citenamefont {Beach}, \citenamefont {Baldo},\ and\ \citenamefont
  {Ross}}]{Currivan2014}%
  \BibitemOpen
  \bibfield  {author} {\bibinfo {author} {\bibfnamefont {J.~A.}\ \bibnamefont
  {Currivan}}, \bibinfo {author} {\bibfnamefont {S.}~\bibnamefont {Siddiqui}},
  \bibinfo {author} {\bibfnamefont {S.}~\bibnamefont {Ahn}}, \bibinfo {author}
  {\bibfnamefont {L.}~\bibnamefont {Tryputen}}, \bibinfo {author}
  {\bibfnamefont {G.~S.~D.}\ \bibnamefont {Beach}}, \bibinfo {author}
  {\bibfnamefont {M.~A.}\ \bibnamefont {Baldo}}, \ and\ \bibinfo {author}
  {\bibfnamefont {C.~A.}\ \bibnamefont {Ross}},\ }\href {\doibase
  10.1116/1.4867753} {\bibfield  {journal} {\bibinfo  {journal} {J. Vac. Sci.
  Technol. B}\ }\textbf {\bibinfo {volume} {32}},\ \bibinfo {pages} {021601}
  (\bibinfo {year} {2014})}\BibitemShut {NoStop}%
\bibitem [{\citenamefont {Nakatani}\ \emph {et~al.}(2005)\citenamefont
  {Nakatani}, \citenamefont {Thiaville},\ and\ \citenamefont
  {Miltat}}]{Nakatani2005}%
  \BibitemOpen
  \bibfield  {author} {\bibinfo {author} {\bibfnamefont {Y.}~\bibnamefont
  {Nakatani}}, \bibinfo {author} {\bibfnamefont {A.}~\bibnamefont {Thiaville}},
  \ and\ \bibinfo {author} {\bibfnamefont {J.}~\bibnamefont {Miltat}},\
  }\href@noop {} {\bibfield  {journal} {\bibinfo  {journal} {J. Magn. Magn.
  Mater.}\ }\textbf {\bibinfo {volume} {290}},\ \bibinfo {pages} {750}
  (\bibinfo {year} {2005})}\BibitemShut {NoStop}%
\bibitem [{\citenamefont {Rhensius}\ \emph {et~al.}(2010)\citenamefont
  {Rhensius}, \citenamefont {Heyne}, \citenamefont {Backes}, \citenamefont
  {Krzyk}, \citenamefont {Heyderman}, \citenamefont {Joly}, \citenamefont
  {Nolting},\ and\ \citenamefont {Klaui}}]{Rhensius2010}%
  \BibitemOpen
  \bibfield  {author} {\bibinfo {author} {\bibfnamefont {J.}~\bibnamefont
  {Rhensius}}, \bibinfo {author} {\bibfnamefont {L.}~\bibnamefont {Heyne}},
  \bibinfo {author} {\bibfnamefont {D.}~\bibnamefont {Backes}}, \bibinfo
  {author} {\bibfnamefont {S.}~\bibnamefont {Krzyk}}, \bibinfo {author}
  {\bibfnamefont {L.~J.}\ \bibnamefont {Heyderman}}, \bibinfo {author}
  {\bibfnamefont {L.}~\bibnamefont {Joly}}, \bibinfo {author} {\bibfnamefont
  {F.}~\bibnamefont {Nolting}}, \ and\ \bibinfo {author} {\bibfnamefont
  {M.}~\bibnamefont {Klaui}},\ }\href@noop {} {\bibfield  {journal} {\bibinfo
  {journal} {Physical Review Letters}\ }\textbf {\bibinfo {volume} {104}},\
  \bibinfo {pages} {067201} (\bibinfo {year} {2010})}\BibitemShut {NoStop}%
\bibitem [{\citenamefont {Wang}\ \emph {et~al.}(2013)\citenamefont {Wang},
  \citenamefont {Guo}, \citenamefont {Bland}, \citenamefont {Nie},
  \citenamefont {Xia},\ and\ \citenamefont {Li}}]{Wang2013b}%
  \BibitemOpen
  \bibfield  {author} {\bibinfo {author} {\bibfnamefont {X.-G.}\ \bibnamefont
  {Wang}}, \bibinfo {author} {\bibfnamefont {G.-H.}\ \bibnamefont {Guo}},
  \bibinfo {author} {\bibfnamefont {J.~A. C.-F.}\ \bibnamefont {Bland}},
  \bibinfo {author} {\bibfnamefont {Y.-Z.}\ \bibnamefont {Nie}}, \bibinfo
  {author} {\bibfnamefont {Q.-L.}\ \bibnamefont {Xia}}, \ and\ \bibinfo
  {author} {\bibfnamefont {Z.-X.}\ \bibnamefont {Li}},\ }\href {\doibase
  10.1016/j.jmmm.2012.12.013} {\bibfield  {journal} {\bibinfo  {journal}
  {Journal of Magnetism and Magnetic Materials}\ }\textbf {\bibinfo {volume}
  {332}},\ \bibinfo {pages} {56} (\bibinfo {year} {2013})}\BibitemShut
  {NoStop}%
\bibitem [{\citenamefont {Stamps}\ \emph {et~al.}(1997)\citenamefont {Stamps},
  \citenamefont {Carri{\c c}o},\ and\ \citenamefont {Wigen}}]{Stamps1997}%
  \BibitemOpen
  \bibfield  {author} {\bibinfo {author} {\bibfnamefont {R.~L.}\ \bibnamefont
  {Stamps}}, \bibinfo {author} {\bibfnamefont {A.~S.}\ \bibnamefont {Carri{\c
  c}o}}, \ and\ \bibinfo {author} {\bibfnamefont {P.~E.}\ \bibnamefont
  {Wigen}},\ }\href@noop {} {\bibfield  {journal} {\bibinfo  {journal} {Phys.
  Rev. Lett.}\ }\textbf {\bibinfo {volume} {55}},\ \bibinfo {pages} {6473}
  (\bibinfo {year} {1997})}\BibitemShut {NoStop}%
\bibitem [{\citenamefont {Liu}\ and\ \citenamefont
  {Gr{\"u}tter}(1998)}]{Liu1998}%
  \BibitemOpen
  \bibfield  {author} {\bibinfo {author} {\bibfnamefont {Y.}~\bibnamefont
  {Liu}}\ and\ \bibinfo {author} {\bibfnamefont {P.}~\bibnamefont
  {Gr{\"u}tter}},\ }\href@noop {} {\bibfield  {journal} {\bibinfo  {journal}
  {J. Appl. Phys.}\ }\textbf {\bibinfo {volume} {83}},\ \bibinfo {pages} {5922}
  (\bibinfo {year} {1998})}\BibitemShut {NoStop}%
\bibitem [{\citenamefont {Dantas}\ \emph {et~al.}(2001)\citenamefont {Dantas},
  \citenamefont {Vasconcelos},\ and\ \citenamefont {Carri{\c
  c}o}}]{Dantas2001}%
  \BibitemOpen
  \bibfield  {author} {\bibinfo {author} {\bibfnamefont {A.~L.}\ \bibnamefont
  {Dantas}}, \bibinfo {author} {\bibfnamefont {M.~S.}\ \bibnamefont
  {Vasconcelos}}, \ and\ \bibinfo {author} {\bibfnamefont {A.~S.}\ \bibnamefont
  {Carri{\c c}o}},\ }\href@noop {} {\bibfield  {journal} {\bibinfo  {journal}
  {J. Magn. Magn. Mater}\ }\textbf {\bibinfo {volume} {226}},\ \bibinfo {pages}
  {1604} (\bibinfo {year} {2001})}\BibitemShut {NoStop}%
\bibitem [{\citenamefont {Matsushita}\ \emph {et~al.}(2012)\citenamefont
  {Matsushita}, \citenamefont {Sasaki}, \citenamefont {Sato},\ and\
  \citenamefont {Imamura}}]{Matsushita2012}%
  \BibitemOpen
  \bibfield  {author} {\bibinfo {author} {\bibfnamefont {K.}~\bibnamefont
  {Matsushita}}, \bibinfo {author} {\bibfnamefont {M.}~\bibnamefont {Sasaki}},
  \bibinfo {author} {\bibfnamefont {J.}~\bibnamefont {Sato}}, \ and\ \bibinfo
  {author} {\bibfnamefont {H.}~\bibnamefont {Imamura}},\ }\href {\doibase
  10.1143/JPSJ.81.043801} {\bibfield  {journal} {\bibinfo  {journal} {J. Phys.
  Soc. Jpn}\ }\textbf {\bibinfo {volume} {81}},\ \bibinfo {pages} {043801}
  (\bibinfo {year} {2012})}\BibitemShut {NoStop}%
\bibitem [{\citenamefont {Mori}\ \emph {et~al.}(2014)\citenamefont {Mori},
  \citenamefont {Koshibae}, \citenamefont {Hikino},\ and\ \citenamefont
  {Maekawa}}]{Mori2014}%
  \BibitemOpen
  \bibfield  {author} {\bibinfo {author} {\bibfnamefont {M.}~\bibnamefont
  {Mori}}, \bibinfo {author} {\bibfnamefont {W.}~\bibnamefont {Koshibae}},
  \bibinfo {author} {\bibfnamefont {S.}~\bibnamefont {Hikino}}, \ and\ \bibinfo
  {author} {\bibfnamefont {S.}~\bibnamefont {Maekawa}},\ }\href@noop {}
  {\bibfield  {journal} {\bibinfo  {journal} {J. Phys.: Condens. Matter}\
  }\textbf {\bibinfo {volume} {26}},\ \bibinfo {pages} {255702} (\bibinfo
  {year} {2014})}\BibitemShut {NoStop}%
\bibitem [{\citenamefont {Matsushita}\ \emph {et~al.}(2014)\citenamefont
  {Matsushita}, \citenamefont {Sasaki},\ and\ \citenamefont
  {Chawanya}}]{Matsushita2014}%
  \BibitemOpen
  \bibfield  {author} {\bibinfo {author} {\bibfnamefont {K.}~\bibnamefont
  {Matsushita}}, \bibinfo {author} {\bibfnamefont {M.}~\bibnamefont {Sasaki}},
  \ and\ \bibinfo {author} {\bibfnamefont {T.}~\bibnamefont {Chawanya}},\
  }\href {\doibase 10.7566/jpsj.83.013801} {\bibfield  {journal} {\bibinfo
  {journal} {J. Phys. Soc. Jpn.}\ }\textbf {\bibinfo {volume} {83}},\ \bibinfo
  {pages} {013801} (\bibinfo {year} {2014})}\BibitemShut {NoStop}%
\bibitem [{\citenamefont {Grimsditch}\ \emph {et~al.}(2004)\citenamefont
  {Grimsditch}, \citenamefont {Leaf}, \citenamefont {Kaper}, \citenamefont
  {Karpeev},\ and\ \citenamefont {Camley}}]{Grimsditch2004}%
  \BibitemOpen
  \bibfield  {author} {\bibinfo {author} {\bibfnamefont {M.}~\bibnamefont
  {Grimsditch}}, \bibinfo {author} {\bibfnamefont {G.~K.}\ \bibnamefont
  {Leaf}}, \bibinfo {author} {\bibfnamefont {H.~G.}\ \bibnamefont {Kaper}},
  \bibinfo {author} {\bibfnamefont {D.~A.}\ \bibnamefont {Karpeev}}, \ and\
  \bibinfo {author} {\bibfnamefont {R.~E.}\ \bibnamefont {Camley}},\
  }\href@noop {} {\bibfield  {journal} {\bibinfo  {journal} {Phys. Rev. B}\
  }\textbf {\bibinfo {volume} {69}},\ \bibinfo {pages} {174428} (\bibinfo
  {year} {2004})}\BibitemShut {NoStop}%
\bibitem [{\citenamefont {McMichael}\ and\ \citenamefont
  {Stiles}(2005)}]{McMichael2005}%
  \BibitemOpen
  \bibfield  {author} {\bibinfo {author} {\bibfnamefont {R.~D.}\ \bibnamefont
  {McMichael}}\ and\ \bibinfo {author} {\bibfnamefont {M.~D.}\ \bibnamefont
  {Stiles}},\ }\href@noop {} {\bibfield  {journal} {\bibinfo  {journal} {J.
  Appl. Phys}\ }\textbf {\bibinfo {volume} {97}},\ \bibinfo {pages} {10J901}
  (\bibinfo {year} {2005})}\BibitemShut {NoStop}%
\bibitem [{\citenamefont {Dvornik}\ \emph {et~al.}(2011)\citenamefont
  {Dvornik}, \citenamefont {Bondarenko}, \citenamefont {Ivanov},\ and\
  \citenamefont {Kruglyak}}]{Dvornik2011}%
  \BibitemOpen
  \bibfield  {author} {\bibinfo {author} {\bibfnamefont {M.}~\bibnamefont
  {Dvornik}}, \bibinfo {author} {\bibfnamefont {P.~V.}\ \bibnamefont
  {Bondarenko}}, \bibinfo {author} {\bibfnamefont {B.~A.}\ \bibnamefont
  {Ivanov}}, \ and\ \bibinfo {author} {\bibfnamefont {V.~V.}\ \bibnamefont
  {Kruglyak}},\ }\href {\doibase 10.1063/1.3562509} {\bibfield  {journal}
  {\bibinfo  {journal} {J. Appl. Phys.}\ }\textbf {\bibinfo {volume} {109}},\
  \bibinfo {pages} {07B912} (\bibinfo {year} {2011})}\BibitemShut {NoStop}%
\bibitem [{\citenamefont {{d'Aquino}}\ \emph {et~al.}(2009)\citenamefont
  {{d'Aquino}}, \citenamefont {Serpico}, \citenamefont {Miano},\ and\
  \citenamefont {Forestiere}}]{dAquino2009}%
  \BibitemOpen
  \bibfield  {author} {\bibinfo {author} {\bibfnamefont {M.}~\bibnamefont
  {{d'Aquino}}}, \bibinfo {author} {\bibfnamefont {C.}~\bibnamefont {Serpico}},
  \bibinfo {author} {\bibfnamefont {G.}~\bibnamefont {Miano}}, \ and\ \bibinfo
  {author} {\bibfnamefont {C.}~\bibnamefont {Forestiere}},\ }\href@noop {}
  {\bibfield  {journal} {\bibinfo  {journal} {J. Comp. Phys.}\ }\textbf
  {\bibinfo {volume} {228}},\ \bibinfo {pages} {6130} (\bibinfo {year}
  {2009})}\BibitemShut {NoStop}%
\bibitem [{\citenamefont {Naletov}\ \emph {et~al.}(2011)\citenamefont
  {Naletov}, \citenamefont {{de Loubens}}, \citenamefont {Albuquerque},
  \citenamefont {Borlenghi}, \citenamefont {Cros}, \citenamefont {Faini},
  \citenamefont {Grollier}, \citenamefont {Hurdequint}, \citenamefont
  {Locatelli}, \citenamefont {Pigeau}, \citenamefont {Slavin}, \citenamefont
  {Tiberkevich}, \citenamefont {Ulysse}, \citenamefont {Valet},\ and\
  \citenamefont {Klein}}]{Naletov2011}%
  \BibitemOpen
  \bibfield  {author} {\bibinfo {author} {\bibfnamefont {V.~V.}\ \bibnamefont
  {Naletov}}, \bibinfo {author} {\bibfnamefont {G.}~\bibnamefont {{de
  Loubens}}}, \bibinfo {author} {\bibfnamefont {G.}~\bibnamefont
  {Albuquerque}}, \bibinfo {author} {\bibfnamefont {S.}~\bibnamefont
  {Borlenghi}}, \bibinfo {author} {\bibfnamefont {V.}~\bibnamefont {Cros}},
  \bibinfo {author} {\bibfnamefont {G.}~\bibnamefont {Faini}}, \bibinfo
  {author} {\bibfnamefont {J.}~\bibnamefont {Grollier}}, \bibinfo {author}
  {\bibfnamefont {H.}~\bibnamefont {Hurdequint}}, \bibinfo {author}
  {\bibfnamefont {N.}~\bibnamefont {Locatelli}}, \bibinfo {author}
  {\bibfnamefont {B.}~\bibnamefont {Pigeau}}, \bibinfo {author} {\bibfnamefont
  {A.~N.}\ \bibnamefont {Slavin}}, \bibinfo {author} {\bibfnamefont {V.~S.}\
  \bibnamefont {Tiberkevich}}, \bibinfo {author} {\bibfnamefont
  {C.}~\bibnamefont {Ulysse}}, \bibinfo {author} {\bibfnamefont
  {T.}~\bibnamefont {Valet}}, \ and\ \bibinfo {author} {\bibfnamefont
  {O.}~\bibnamefont {Klein}},\ }\href@noop {} {\bibfield  {journal} {\bibinfo
  {journal} {Phys. Rev. B}\ }\textbf {\bibinfo {volume} {84}},\ \bibinfo
  {pages} {224423} (\bibinfo {year} {2011})}\BibitemShut {NoStop}%
\bibitem [{\citenamefont {Zivieri}\ and\ \citenamefont
  {Consolo}(2012)}]{Zivieri2012a}%
  \BibitemOpen
  \bibfield  {author} {\bibinfo {author} {\bibfnamefont {R.}~\bibnamefont
  {Zivieri}}\ and\ \bibinfo {author} {\bibfnamefont {G.}~\bibnamefont
  {Consolo}},\ }\href {\doibase 10.1155/2012/765709} {\bibfield  {journal}
  {\bibinfo  {journal} {Adv. Cond. Matt. Phys.}\ }\textbf {\bibinfo {volume}
  {2012}},\ \bibinfo {pages} {1} (\bibinfo {year} {2012})}\BibitemShut
  {NoStop}%
\bibitem [{\citenamefont {Fischbacher}\ \emph {et~al.}(2007)\citenamefont
  {Fischbacher}, \citenamefont {Franchin}, \citenamefont {Bordignon},\ and\
  \citenamefont {Fangohr}}]{Fischbacher2007}%
  \BibitemOpen
  \bibfield  {author} {\bibinfo {author} {\bibfnamefont {T.}~\bibnamefont
  {Fischbacher}}, \bibinfo {author} {\bibfnamefont {M.}~\bibnamefont
  {Franchin}}, \bibinfo {author} {\bibfnamefont {G.}~\bibnamefont {Bordignon}},
  \ and\ \bibinfo {author} {\bibfnamefont {H.}~\bibnamefont {Fangohr}},\
  }\href@noop {} {\bibfield  {journal} {\bibinfo  {journal} {IEEE Trans. Mag.}\
  }\textbf {\bibinfo {volume} {43}},\ \bibinfo {pages} {2896} (\bibinfo {year}
  {2007})}\BibitemShut {NoStop}%
\bibitem [{Note1()}]{Note1}%
  \BibitemOpen
  \bibinfo {note} {The authors intend to include supplementary animations with the final published version of the paper.}\BibitemShut {Stop}%
\bibitem [{\citenamefont {Abo}\ \emph {et~al.}(2013)\citenamefont {Abo},
  \citenamefont {Hong}, \citenamefont {Park}, \citenamefont {Lee},
  \citenamefont {Lee},\ and\ \citenamefont {Choi}}]{Abo2013}%
  \BibitemOpen
  \bibfield  {author} {\bibinfo {author} {\bibfnamefont {G.~S.}\ \bibnamefont
  {Abo}}, \bibinfo {author} {\bibfnamefont {Y.-K.}\ \bibnamefont {Hong}},
  \bibinfo {author} {\bibfnamefont {J.-H.}\ \bibnamefont {Park}}, \bibinfo
  {author} {\bibfnamefont {J.-J.}\ \bibnamefont {Lee}}, \bibinfo {author}
  {\bibfnamefont {W.}~\bibnamefont {Lee}}, \ and\ \bibinfo {author}
  {\bibfnamefont {B.-C.}\ \bibnamefont {Choi}},\ }\href@noop {} {\bibfield
  {journal} {\bibinfo  {journal} {IEEE Trans. Mag.}\ }\textbf {\bibinfo
  {volume} {49}},\ \bibinfo {pages} {4937} (\bibinfo {year}
  {2013})}\BibitemShut {NoStop}%
\bibitem [{\citenamefont {Metaxas}\ \emph {et~al.}(2015)\citenamefont
  {Metaxas}, \citenamefont {Sushruth}, \citenamefont {Begley}, \citenamefont
  {Ding}, \citenamefont {Woodward}, \citenamefont {Maksymov}, \citenamefont
  {Albert}, \citenamefont {Wang}, \citenamefont {Fangohr}, \citenamefont
  {Adeyeye},\ and\ \citenamefont {Kostylev}}]{Metaxas2015}%
  \BibitemOpen
  \bibfield  {author} {\bibinfo {author} {\bibfnamefont {P.~J.}\ \bibnamefont
  {Metaxas}}, \bibinfo {author} {\bibfnamefont {M.}~\bibnamefont {Sushruth}},
  \bibinfo {author} {\bibfnamefont {R.}~\bibnamefont {Begley}}, \bibinfo
  {author} {\bibfnamefont {J.}~\bibnamefont {Ding}}, \bibinfo {author}
  {\bibfnamefont {R.~C.}\ \bibnamefont {Woodward}}, \bibinfo {author}
  {\bibfnamefont {I.}~\bibnamefont {Maksymov}}, \bibinfo {author}
  {\bibfnamefont {M.}~\bibnamefont {Albert}}, \bibinfo {author} {\bibfnamefont
  {W.}~\bibnamefont {Wang}}, \bibinfo {author} {\bibfnamefont {H.}~\bibnamefont
  {Fangohr}}, \bibinfo {author} {\bibfnamefont {A.}~\bibnamefont {Adeyeye}}, \
  and\ \bibinfo {author} {\bibfnamefont {M.}~\bibnamefont {Kostylev}},\
  }\href@noop {} {\bibfield  {journal} {\bibinfo  {journal} {Appl. Phys.
  Lett.}\ }\textbf {\bibinfo {volume} {106}},\ \bibinfo {pages} {232406}
  (\bibinfo {year} {2015})}\BibitemShut {NoStop}%
\bibitem [{\citenamefont {Martinez}\ \emph {et~al.}(2007)\citenamefont
  {Martinez}, \citenamefont {{Lopez-Diaz}}, \citenamefont {Alejos},
  \citenamefont {Torres},\ and\ \citenamefont {Tristan}}]{Martinez2007a}%
  \BibitemOpen
  \bibfield  {author} {\bibinfo {author} {\bibfnamefont {E.}~\bibnamefont
  {Martinez}}, \bibinfo {author} {\bibfnamefont {L.}~\bibnamefont
  {{Lopez-Diaz}}}, \bibinfo {author} {\bibfnamefont {O.}~\bibnamefont
  {Alejos}}, \bibinfo {author} {\bibfnamefont {L.}~\bibnamefont {Torres}}, \
  and\ \bibinfo {author} {\bibfnamefont {C.}~\bibnamefont {Tristan}},\
  }\href@noop {} {\bibfield  {journal} {\bibinfo  {journal} {Phys. Rev. Lett.}\
  }\textbf {\bibinfo {volume} {98}},\ \bibinfo {pages} {267202} (\bibinfo
  {year} {2007})}\BibitemShut {NoStop}%
\bibitem [{\citenamefont {Tatara}\ and\ \citenamefont
  {Kohno}(2004)}]{Tatara2004}%
  \BibitemOpen
  \bibfield  {author} {\bibinfo {author} {\bibfnamefont {G.}~\bibnamefont
  {Tatara}}\ and\ \bibinfo {author} {\bibfnamefont {H.}~\bibnamefont {Kohno}},\
  }\href@noop {} {\bibfield  {journal} {\bibinfo  {journal} {Physical Review
  Letters}\ }\textbf {\bibinfo {volume} {92}},\ \bibinfo {pages} {086601}
  (\bibinfo {year} {2004})}\BibitemShut {NoStop}%
\bibitem [{\citenamefont {Kr\"uger}(2011)}]{KrugerThesis}%
  \BibitemOpen
  \bibfield  {author} {\bibinfo {author} {\bibfnamefont {B.}~\bibnamefont
  {Kr\"uger}},\ }\emph {\bibinfo {title} {Current-Driven Magnetization
  Dynamics: Analytical Modeling and Numerical Simulation}},\ \href@noop {}
  {Ph.D. thesis},\ \bibinfo  {school} {Universit\"at Hamburg} (\bibinfo {year}
  {2011})\BibitemShut {NoStop}%
\bibitem [{\citenamefont {Thiaville}\ \emph {et~al.}(2007)\citenamefont
  {Thiaville}, \citenamefont {Nakatani}, \citenamefont {Piￃﾩchon},
  \citenamefont {Miltat},\ and\ \citenamefont {Ono}}]{Thiaville2007}%
  \BibitemOpen
  \bibfield  {author} {\bibinfo {author} {\bibfnamefont {A.}~\bibnamefont
  {Thiaville}}, \bibinfo {author} {\bibfnamefont {Y.}~\bibnamefont {Nakatani}},
  \bibinfo {author} {\bibfnamefont {F.}~\bibnamefont {Piￃﾩchon}}, \bibinfo
  {author} {\bibfnamefont {J.}~\bibnamefont {Miltat}}, \ and\ \bibinfo {author}
  {\bibfnamefont {T.}~\bibnamefont {Ono}},\ }\href {\doibase
  10.1140/epjb/e2007-00320-3} {\bibfield  {journal} {\bibinfo  {journal} {Eur.
  Phys. J. B}\ }\textbf {\bibinfo {volume} {60}},\ \bibinfo {pages}
  {15￢ﾀﾓ27} (\bibinfo {year} {2007})}\BibitemShut {NoStop}%
\bibitem [{\citenamefont {Aharoni}(1998)}]{Aharoni1998}%
  \BibitemOpen
  \bibfield  {author} {\bibinfo {author} {\bibfnamefont {A.}~\bibnamefont
  {Aharoni}},\ }\href@noop {} {\bibfield  {journal} {\bibinfo  {journal} {J.
  Appl. Phys.}\ }\textbf {\bibinfo {volume} {83}},\ \bibinfo {pages} {3432}
  (\bibinfo {year} {1998})}\BibitemShut {NoStop}%
\bibitem [{\citenamefont {Moriya}\ \emph {et~al.}(2008)\citenamefont {Moriya},
  \citenamefont {Thomas}, \citenamefont {Hayashi}, \citenamefont {Bazaliy},
  \citenamefont {Rettner},\ and\ \citenamefont {Parkin}}]{Moriya2008}%
  \BibitemOpen
  \bibfield  {author} {\bibinfo {author} {\bibfnamefont {R.}~\bibnamefont
  {Moriya}}, \bibinfo {author} {\bibfnamefont {L.}~\bibnamefont {Thomas}},
  \bibinfo {author} {\bibfnamefont {M.}~\bibnamefont {Hayashi}}, \bibinfo
  {author} {\bibfnamefont {Y.~B.}\ \bibnamefont {Bazaliy}}, \bibinfo {author}
  {\bibfnamefont {C.}~\bibnamefont {Rettner}}, \ and\ \bibinfo {author}
  {\bibfnamefont {S.~S.~P.}\ \bibnamefont {Parkin}},\ }\href {\doibase
  10.1038/nphys936} {\bibfield  {journal} {\bibinfo  {journal} {Nat Phys}\
  }\textbf {\bibinfo {volume} {4}},\ \bibinfo {pages} {368￢ﾀﾓ372}
  (\bibinfo {year} {2008})}\BibitemShut {NoStop}%
\bibitem [{\citenamefont {Kim}\ \emph {et~al.}(2014)\citenamefont {Kim},
  \citenamefont {Mawass}, \citenamefont {Bisig}, \citenamefont {Kr{\"u}ger},
  \citenamefont {Reeve}, \citenamefont {Schulz}, \citenamefont {B{\"u}ttner},
  \citenamefont {Yoon}, \citenamefont {You}, \citenamefont {Weigand},\ and\
  \citenamefont {et~al.}}]{Kim2014}%
  \BibitemOpen
  \bibfield  {author} {\bibinfo {author} {\bibfnamefont {J.-S.}\ \bibnamefont
  {Kim}}, \bibinfo {author} {\bibfnamefont {M.-A.}\ \bibnamefont {Mawass}},
  \bibinfo {author} {\bibfnamefont {A.}~\bibnamefont {Bisig}}, \bibinfo
  {author} {\bibfnamefont {B.}~\bibnamefont {Kr{\"u}ger}}, \bibinfo {author}
  {\bibfnamefont {R.~M.}\ \bibnamefont {Reeve}}, \bibinfo {author}
  {\bibfnamefont {T.}~\bibnamefont {Schulz}}, \bibinfo {author} {\bibfnamefont
  {F.}~\bibnamefont {B{\"u}ttner}}, \bibinfo {author} {\bibfnamefont
  {J.}~\bibnamefont {Yoon}}, \bibinfo {author} {\bibfnamefont {C.-Y.}\
  \bibnamefont {You}}, \bibinfo {author} {\bibfnamefont {M.}~\bibnamefont
  {Weigand}}, \ and\ \bibinfo {author} {\bibnamefont {et~al.}},\ }\href@noop {}
  {\bibfield  {journal} {\bibinfo  {journal} {Nat. Commun.}\ }\textbf {\bibinfo
  {volume} {5}} (\bibinfo {year} {2014})}\BibitemShut {NoStop}%
\bibitem [{\citenamefont {Bocklage}\ \emph {et~al.}(2014)\citenamefont
  {Bocklage}, \citenamefont {Motl-Ziegler}, \citenamefont {Topp}, \citenamefont
  {Matsuyama},\ and\ \citenamefont {Meier}}]{Bocklage2014}%
  \BibitemOpen
  \bibfield  {author} {\bibinfo {author} {\bibfnamefont {L.}~\bibnamefont
  {Bocklage}}, \bibinfo {author} {\bibfnamefont {S.}~\bibnamefont
  {Motl-Ziegler}}, \bibinfo {author} {\bibfnamefont {J.}~\bibnamefont {Topp}},
  \bibinfo {author} {\bibfnamefont {T.}~\bibnamefont {Matsuyama}}, \ and\
  \bibinfo {author} {\bibfnamefont {G.}~\bibnamefont {Meier}},\ }\href
  {\doibase 10.1088/0953-8984/26/26/266003} {\bibfield  {journal} {\bibinfo
  {journal} {J. Phys.: Condens. Matter}\ }\textbf {\bibinfo {volume} {26}},\
  \bibinfo {pages} {266003} (\bibinfo {year} {2014})}\BibitemShut {NoStop}%
\bibitem [{\citenamefont {Uhlir}\ \emph {et~al.}(2011)\citenamefont {Uhlir},
  \citenamefont {{Pizzini}}, \citenamefont {{Rougemaille}}, \citenamefont
  {{Cros}}, \citenamefont {{Jimenez}}, \citenamefont {{Ranno}}, \citenamefont
  {{Fruchart}}, \citenamefont {{Urbanek}}, \citenamefont {{Gaudin}},
  \citenamefont {{Camarero}}, \citenamefont {{Tieg}}, \citenamefont
  {{Sirotti}}, \citenamefont {Wagner},\ and\ \citenamefont
  {{Vogel}}}]{Uhlir2011}%
  \BibitemOpen
  \bibfield  {author} {\bibinfo {author} {\bibfnamefont {V.}~\bibnamefont
  {Uhlir}}, \bibinfo {author} {\bibfnamefont {S.}~\bibnamefont {{Pizzini}}},
  \bibinfo {author} {\bibfnamefont {N.}~\bibnamefont {{Rougemaille}}}, \bibinfo
  {author} {\bibfnamefont {V.}~\bibnamefont {{Cros}}}, \bibinfo {author}
  {\bibfnamefont {E.}~\bibnamefont {{Jimenez}}}, \bibinfo {author}
  {\bibfnamefont {L.}~\bibnamefont {{Ranno}}}, \bibinfo {author} {\bibfnamefont
  {O.}~\bibnamefont {{Fruchart}}}, \bibinfo {author} {\bibfnamefont
  {M.}~\bibnamefont {{Urbanek}}}, \bibinfo {author} {\bibfnamefont
  {G.}~\bibnamefont {{Gaudin}}}, \bibinfo {author} {\bibfnamefont
  {J.}~\bibnamefont {{Camarero}}}, \bibinfo {author} {\bibfnamefont
  {C.}~\bibnamefont {{Tieg}}}, \bibinfo {author} {\bibfnamefont
  {F.}~\bibnamefont {{Sirotti}}}, \bibinfo {author} {\bibfnamefont
  {E.}~\bibnamefont {Wagner}}, \ and\ \bibinfo {author} {\bibfnamefont
  {J.}~\bibnamefont {{Vogel}}},\ }\href@noop {} {\bibfield  {journal} {\bibinfo
   {journal} {Phys. Rev. B}\ }\textbf {\bibinfo {volume} {83}},\ \bibinfo
  {pages} {020406R} (\bibinfo {year} {2011})}\BibitemShut {NoStop}%
\bibitem [{\citenamefont {Chanthbouala}\ \emph {et~al.}(2011)\citenamefont
  {Chanthbouala}, \citenamefont {Matsumoto}, \citenamefont {Grollier},
  \citenamefont {Cros}, \citenamefont {Anane}, \citenamefont {Fert},
  \citenamefont {Khvalkovskiy}, \citenamefont {Zvezdin}, \citenamefont
  {Nishimura}, \citenamefont {Nagamine}, \citenamefont {Maehara}, \citenamefont
  {Tsunekawa}, \citenamefont {Fukushima},\ and\ \citenamefont
  {Yuasa}}]{Chanthbouala2011}%
  \BibitemOpen
  \bibfield  {author} {\bibinfo {author} {\bibfnamefont {A.}~\bibnamefont
  {Chanthbouala}}, \bibinfo {author} {\bibfnamefont {R.}~\bibnamefont
  {Matsumoto}}, \bibinfo {author} {\bibfnamefont {J.}~\bibnamefont {Grollier}},
  \bibinfo {author} {\bibfnamefont {V.}~\bibnamefont {Cros}}, \bibinfo {author}
  {\bibfnamefont {A.}~\bibnamefont {Anane}}, \bibinfo {author} {\bibfnamefont
  {A.}~\bibnamefont {Fert}}, \bibinfo {author} {\bibfnamefont {A.~V.}\
  \bibnamefont {Khvalkovskiy}}, \bibinfo {author} {\bibfnamefont {K.~A.}\
  \bibnamefont {Zvezdin}}, \bibinfo {author} {\bibfnamefont {N.}~\bibnamefont
  {Nishimura}}, \bibinfo {author} {\bibfnamefont {Y.}~\bibnamefont {Nagamine}},
  \bibinfo {author} {\bibfnamefont {H.}~\bibnamefont {Maehara}}, \bibinfo
  {author} {\bibfnamefont {K.}~\bibnamefont {Tsunekawa}}, \bibinfo {author}
  {\bibfnamefont {A.}~\bibnamefont {Fukushima}}, \ and\ \bibinfo {author}
  {\bibfnamefont {S.}~\bibnamefont {Yuasa}},\ }\href@noop {} {\bibfield
  {journal} {\bibinfo  {journal} {Nat. Phys.}\ }\textbf {\bibinfo {volume}
  {7}},\ \bibinfo {pages} {626} (\bibinfo {year} {2011})}\BibitemShut {NoStop}%
\bibitem [{\citenamefont {Khvalkovskiy}\ \emph {et~al.}(2009)\citenamefont
  {Khvalkovskiy}, \citenamefont {Zvezdin}, \citenamefont {Gorbunov},
  \citenamefont {Cros}, \citenamefont {Grollier}, \citenamefont {Fert},\ and\
  \citenamefont {Zvezdin}}]{Khvalkovskiy2009}%
  \BibitemOpen
  \bibfield  {author} {\bibinfo {author} {\bibfnamefont {A.}~\bibnamefont
  {Khvalkovskiy}}, \bibinfo {author} {\bibfnamefont {K.~A.}\ \bibnamefont
  {Zvezdin}}, \bibinfo {author} {\bibfnamefont {Y.~V.}\ \bibnamefont
  {Gorbunov}}, \bibinfo {author} {\bibfnamefont {V.}~\bibnamefont {Cros}},
  \bibinfo {author} {\bibfnamefont {J.}~\bibnamefont {Grollier}}, \bibinfo
  {author} {\bibfnamefont {A.}~\bibnamefont {Fert}}, \ and\ \bibinfo {author}
  {\bibfnamefont {A.~K.}\ \bibnamefont {Zvezdin}},\ }\href@noop {} {\bibfield
  {journal} {\bibinfo  {journal} {Phys. Rev. Lett.}\ }\textbf {\bibinfo
  {volume} {102}},\ \bibinfo {pages} {067206} (\bibinfo {year}
  {2009})}\BibitemShut {NoStop}%
\bibitem [{\citenamefont {Boone}\ \emph {et~al.}(2010)\citenamefont {Boone},
  \citenamefont {Katine}, \citenamefont {Carey}, \citenamefont {Childress},
  \citenamefont {Cheng},\ and\ \citenamefont {Krivorotov}}]{Boone2010}%
  \BibitemOpen
  \bibfield  {author} {\bibinfo {author} {\bibfnamefont {C.~T.}\ \bibnamefont
  {Boone}}, \bibinfo {author} {\bibfnamefont {J.~A.}\ \bibnamefont {Katine}},
  \bibinfo {author} {\bibfnamefont {M.}~\bibnamefont {Carey}}, \bibinfo
  {author} {\bibfnamefont {J.~R.}\ \bibnamefont {Childress}}, \bibinfo {author}
  {\bibfnamefont {X.}~\bibnamefont {Cheng}}, \ and\ \bibinfo {author}
  {\bibfnamefont {I.N.}~\bibnamefont {Krivorotov}},\ }\href@noop {} {\bibfield
  {journal} {\bibinfo  {journal} {Phys. Rev. Lett.}\ }\textbf {\bibinfo
  {volume} {104}},\ \bibinfo {pages} {097203} (\bibinfo {year}
  {2010})}\BibitemShut {NoStop}%
\bibitem [{\citenamefont {Metaxas}\ \emph {et~al.}(2013)\citenamefont
  {Metaxas}, \citenamefont {Sampaio}, \citenamefont {Chanthbouala},
  \citenamefont {Matsumoto}, \citenamefont {Anane}, \citenamefont {Zvezdin},
  \citenamefont {Yakushiji}, \citenamefont {Kubota}, \citenamefont {Fukushima},
  \citenamefont {Yuasa}, \citenamefont {Nishimura}, \citenamefont {Nagamine},
  \citenamefont {Maehara}, \citenamefont {Tsunekawa}, \citenamefont {Cros},\
  and\ \citenamefont {Grollier}}]{Metaxas2013b}%
  \BibitemOpen
  \bibfield  {author} {\bibinfo {author} {\bibfnamefont {P.~J.}\ \bibnamefont
  {Metaxas}}, \bibinfo {author} {\bibfnamefont {J.}~\bibnamefont {Sampaio}},
  \bibinfo {author} {\bibfnamefont {A.}~\bibnamefont {Chanthbouala}}, \bibinfo
  {author} {\bibfnamefont {R.}~\bibnamefont {Matsumoto}}, \bibinfo {author}
  {\bibfnamefont {A.}~\bibnamefont {Anane}}, \bibinfo {author} {\bibfnamefont
  {A.~K.}\ \bibnamefont {Zvezdin}}, \bibinfo {author} {\bibfnamefont
  {K.}~\bibnamefont {Yakushiji}}, \bibinfo {author} {\bibfnamefont
  {H.}~\bibnamefont {Kubota}}, \bibinfo {author} {\bibfnamefont
  {A.}~\bibnamefont {Fukushima}}, \bibinfo {author} {\bibfnamefont
  {S.}~\bibnamefont {Yuasa}}, \bibinfo {author} {\bibfnamefont
  {K.}~\bibnamefont {Nishimura}}, \bibinfo {author} {\bibfnamefont
  {Y.}~\bibnamefont {Nagamine}}, \bibinfo {author} {\bibfnamefont
  {H.}~\bibnamefont {Maehara}}, \bibinfo {author} {\bibfnamefont
  {K.}~\bibnamefont {Tsunekawa}}, \bibinfo {author} {\bibfnamefont
  {V.}~\bibnamefont {Cros}}, \ and\ \bibinfo {author} {\bibfnamefont
  {J.}~\bibnamefont {Grollier}},\ }\href@noop {} {\bibfield  {journal}
  {\bibinfo  {journal} {Sci. Rep.}\ }\textbf {\bibinfo {volume} {3}},\ \bibinfo
  {pages} {1829} (\bibinfo {year} {2013})}\BibitemShut {NoStop}%
\bibitem [{\citenamefont {Bryan}\ \emph {et~al.}(2012)\citenamefont {Bryan},
  \citenamefont {Bance}, \citenamefont {Dean}, \citenamefont {Schrefl},\ and\
  \citenamefont {Allwood}}]{Bryan2012}%
  \BibitemOpen
  \bibfield  {author} {\bibinfo {author} {\bibfnamefont {M.~T.}\ \bibnamefont
  {Bryan}}, \bibinfo {author} {\bibfnamefont {S.}~\bibnamefont {Bance}},
  \bibinfo {author} {\bibfnamefont {J.}~\bibnamefont {Dean}}, \bibinfo {author}
  {\bibfnamefont {T.}~\bibnamefont {Schrefl}}, \ and\ \bibinfo {author}
  {\bibfnamefont {D.~A.}\ \bibnamefont {Allwood}},\ }\href {\doibase
  10.1088/0953-8984/24/2/024205} {\bibfield  {journal} {\bibinfo  {journal} {J.
  Phys.: Condens. Matter}\ }\textbf {\bibinfo {volume} {24}},\ \bibinfo {pages}
  {024205} (\bibinfo {year} {2012})}\BibitemShut {NoStop}%
\end{thebibliography}

%

\end{document}